\documentclass{emulateapj}

\usepackage{lscape}

\makeatother

\newcommand{\npropto}{\ \raise -0.truept\hbox{\rlap{\hbox{$/$}}\raise0.truept
        \hbox{$\propto$}\ }}

\newcommand{\lsim}{\ \raise -2.truept\hbox{\rlap{\hbox{$\sim$}}\raise5.truept
        \hbox{$<$}\ }}
\newcommand{\gsim}{\ \raise -2.truept\hbox{\rlap{\hbox{$\sim$}}\raise5.truept
        \hbox{$>$}\ }} 

\newcommand{\cm}{cm$^{-2}$}

\slugcomment{ApJ in press}
\shorttitle{GDDS: Redshift Evolution of Mass-Metallicity Relation}
\shortauthors{Savaglio, Glazebrook, Le Borgne et al.}

\begin{document}

\title{THE GEMINI DEEP DEEP SURVEY. VII. The Redshift Evolution of the 
Mass-Metallicity Relation$^{1,2}$}

\author{S.\ Savaglio\altaffilmark{3}, K.\
Glazebrook\altaffilmark{3}, D.\ Le Borgne\altaffilmark{4},
S.\ Juneau\altaffilmark{5,6}, R.\ G.\
Abraham\altaffilmark{4}, H.-W.\ Chen\altaffilmark{7,11}, D.\
Crampton\altaffilmark{6}, P.\ J.\ McCarthy\altaffilmark{8},
R.\ G.\ Carlberg\altaffilmark{4}, R.\ O.\
Marzke\altaffilmark{9}, K.\ Roth\altaffilmark{10}, I.\
J{\o}rgensen\altaffilmark{10}, R.\ Murowinski\altaffilmark{6}}

\altaffiltext{1}{Based on observations obtained at the Gemini
Observatory}

\altaffiltext{2}{Based on observations obtained at the
Canada-France-Hawaii Telescope}

\altaffiltext{3}{Department of Physics \& Astronomy, Johns Hopkins
University, Baltimore, MD 21218, [kgb; savaglio]@pha.jhu.edu}

\altaffiltext{4}{Department of Astronomy \& Astrophysics, University
of Toronto, Toronto ON, M5S~3H8 Canada, [leborgne; abraham;
carlberg]@astro.utoronto.ca}

\altaffiltext{5}{D\'{e}partement de physique, Universit\'{e} de
Montr\'eal, 29 00, Bld. \'{E}douard-Montpetit, Montr\'{e}al, QC,
Canada H3T 1J4, sjuneau@astro.umontreal.ca}

\altaffiltext{6}{NRC Herzberg Institute for Astrophysics, 5071
W. Saanich Rd., Victoria, BC, Canada, [david.crampton;
murowinski]@nrc-cnrc.gc.ca}

\altaffiltext{7}{Center for Space Sciences, Massachusetts Institute of
Technology, 70 Vassar St., Bld. 37, Cambridge, MA 02139,
hchen@space.mit.edu}

\altaffiltext{8}{Carnegie Observatories, 813 Santa Barbara St,
Pasadena, CA 91101, pmc2@ociw.edu}

\altaffiltext{9}{Department of Physics and Astronomy, San Francisco
State University, San Francisco, CA 94132, marzke@stars.sfsu.edu}

\altaffiltext{10}{Gemini Observatory, 670 North A'ohoku Place, Hilo, HI
97620, [jorgensen; kroth]@gemini.edu}

\altaffiltext{11}{Hubble Fellow}

\begin{abstract} 

We have investigated the mass-metallicity (M-Z) relation using
galaxies at $0.4<z<1.0$ from the Gemini Deep Deep Survey (GDDS) and
Canada-France Redshift Survey (CFRS).  Deep $K$ and $z'$ band
photometry allowed us to measure stellar masses for 69 galaxies. From
a subsample of 56 galaxies, for which metallicity of the interstellar
medium is also measured, we identified a strong
correlation between mass and metallicity, for the first time in the
distant Universe. This was possible because of the larger base line
spanned by the sample in terms of metallicity (a factor of 7) and mass
(a factor of 400) than in previous works. This correlation is much
stronger and tighter than the luminosity-metallicity, confirming that
stellar mass is a more meaningful physical parameter than luminosity.
We find clear evidence for temporal evolution in the M-Z relation in
the sense that at a given mass, a galaxy at $z\sim0.7$ tends to have
lower metallicity than a local galaxy of similar mass. We use the
$z\sim0.1$ Sloan Digital Sky Survey M-Z relation, and a small sample
of $z\sim2.3$ Lyman break galaxies with known mass and metallicity, to
propose an empirical redshift-dependent M-Z relation, according to
which the stellar mass and metallicity in small galaxies evolve for a
longer time than in massive galaxies. This relation predicts that
the generally metal poor damped Lyman-$\alpha$ galaxies have
stellar masses of the order of $10^{8.8}$ M$_\odot$ (with a dispersion
of 0.7 dex) all the way from $z\sim0.2$ to $z\sim4$. The observed
redshift evolution of the M-Z relation can be reproduced remarkably
well by a simple closed-box model where the key assumption is an
e-folding time for star formation which is higher or, in other words,
a period of star formation that lasts longer in less massive galaxies
than in more massive galaxies. Such a picture supports the downsizing
scenario for galaxy formation.
\end{abstract}

\keywords{galaxies: fundamental parameters -- galaxies: abundances --
galaxies: ISM -- ISM: H II regions -- galaxies: evolution --
cosmology: observations}

\section{Introduction}

Our exploration of the evolution of the cosmic metal enrichment relies
mainly on two methods. One is based on the detection of absorption
lines in the neutral interstellar medium (ISM) of galaxies crossing
QSO sight-lines (Prochaska et al. 2003), and gives information of one
line of sight in the galaxy.  The other uses emission lines of the
warm ISM (HII regions) detected in the integrated galaxy spectra, and
is observationally much more challenging at high redshift because
important lines get very weak and redshifted to the near
IR.\footnote{Stellar features can also be used as metallicity
estimator, however this way it is much harder to obtain accurate
information for a single galaxy (Pettini et al.\ 2000; de Mello et
al.\ 2004).}

The warm ISM metallicity and the mass of galaxies are strongly
correlated in the low-$z$ Universe (Lequeux et al.\ 1979). More
massive galaxies have higher metallicity than less massive galaxies.
This mass-metallicity (M-Z) relation has never been detected at high
redshift. In fact, measurements of the stellar mass (strongly
correlated with the dynamical mass; Brinchmann \& Ellis 2000) require
deep optical/NIR photometry of faint targets, and are not easy to
obtain for a sufficiently large sample for which metallicity is known.

To explore any evolution with redshift, luminosity, as a proxy for
mass, can be used instead, although luminosity is more difficult to
interpret physically than mass.  The metallicity of large $0.3<z<1$
galaxy samples has been studied in relation to the galaxy luminosity
by Kobulnicky et al.\ (2003), Lilly, Carollo, \& Stockton (2003),
Kobulnicky \& Kewley (2004) and Liang et al.\ (2004). Kobulnicky \&
Kewley (2004; hereafter KK04) have used about 200 galaxies from the
Team Keck Redshift Survey (TKRS; Wirth et al.\ 2004) and found a
luminosity-metallicity (L-Z) relation displaced towards higher
luminosities with respect to the same relation detected at $z=0$
(Kennicutt 1992; Jansen et al.\ 2000).  This suggests a redshift
evolution of the L-Z relation.  At higher $z$, only a handful of
galaxies have been studied so far (Pettini et al.\ 2001; Shapley et
al. 2004) and again low metallicities are found compared to galaxies
with similar luminosity at low redshifts.

Do we also expect the M-Z relation to evolve with time? Shifts of
the L-Z relation will arise from simple luminosity evolution
(i.e. changing the mass--to-light ratio) and this is well documented in
galaxies over $0<z<1$ (Schade et al.\ 1996; Vogt et al.\ 1996). In
particular passive fading of stellar populations (as observed in
early-type galaxies, e.g. Aragon-Salamanca et al.\ 1993) would shift
L-Z but not M-Z. However, astrophysically, we would expect the
star-formation activity over $0<z<1$ to manifest itself in changes of
metallicity as well as mass. The question is what is the form of this
change and can it be detected?

From a different prospective, the cold ISM, damped Lyman-$\alpha$
(DLA) galaxies detected in QSO spectra have metallicities $\sim1/10$
solar at $z\sim0.7$ (Prochaska et al.\ 2003), i.e.\ one order of
magnitudes lower than line emitters at similar redshifts (KK04).  Part
of the discrepancy could be due to the fact that the absorption lines
statistically probe the outskirt of galaxies, where metallicity is
lower, while emission line flux originates in the central more metal
rich region (Chen, Kennicutt, \& Rauch 2005; Ellison et al. 2005;
Schulte-Ladbeck et al.\ 2005). Another possibility is that DLAs and
line emitters are probing different galaxies all together.
Unfortunately, so far no information on masses of DLA galaxies is
available.

In this work we present the first attempt to derive the
mass-metallicity relation in high redshift galaxies selected by
near-IR photometry down to small masses. The sample is selected from
the $0.4<z<1$ galaxies of the Gemini Deep Deep Survey (GDDS; Abraham
et al. 2004) and Canada-France Redshift Survey (CFRS; Crampton et al.\
1995; Le F{\`e}vre et al.\ 1995; Lilly et al.\ 1995). Masses were
estimated using deep optical-NIR photometry and metallicities using
optical spectroscopy. For consistency with GDDS, metallicities for the
CFRS galaxies were recomputed using emission line fluxes published by
Lilly et al.\ (2003; hereafter LCS).

Before continuing our discussion it is important to emphasize that
measurements of physical parameters can differ by some factor,
depending on how they are estimated. For instance, masses can change
by a factor of $\sim2$, depending on which initial mass function (IMF)
is applied. Similarly, metallicities can be different by a factor of
$2-3$, depending on which set of lines and/or calibrator are
considered (Kennicutt, Bresolin \& Garnett 2003; KK04). Thus, we paid
special attention to deriving results, and comparing them with other
works, in a consistent fashion.

The paper is organized as follows: in \S2, we describe the sample
section from the GDDS; in \S3 we present the GDDS composite spectrum
used to derive an overall stellar absorption and dust extinction; \S4
\& \S5 describe the stellar mass and the metallicity derivations,
respectively, for the GDDS and the CFRS samples; \S6 \& \S7 are about
the luminosity-metallicity and mass-metallicity relations,
respectively; in \S8 we discuss possible systematic effects; \S9 is on
the modeling of the redshift evolution of the M-Z relation; the
discussion and the concluding remarks are in \S10 and \S11.
Throughout the paper we adopt a $h \equiv H_o/100= 0.7$, $\Omega_M =
0.3$, $\Omega_\Lambda = 0.7$ cosmology (Spergel et al.\ 2003).

\begin{figure}
\centerline{\epsfxsize=8.5cm \epsffile{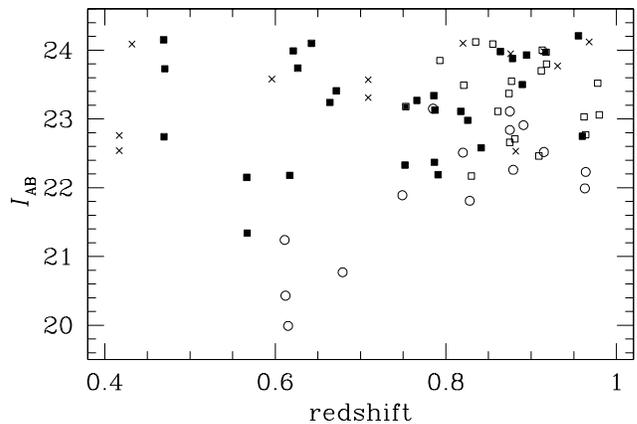}} 
\caption {The $I_{AB}$ magnitude vs.  redshift of the GDDS galaxies at
$0.40<z<0.98$. The selected galaxies are shown as filled squares; 25\%
and 75\% are classified as late-type and intermediate-type galaxies,
respectively.  For the rest of the sample, pure early-type and
intermediate-type galaxies are shown as empty circles and squares,
respectively.  The non-selected intermediate-type galaxies, generally
show only [OII] emission and no other line. One galaxy (SA22-2107) is
not included in the selected sample because it is contaminated by an
interloper. Crosses are galaxies not included in the sample because
one or more emission lines are corrupted or outside the spectral
range.}
\label{f1}
\end{figure}

\section{The GDDS sample selection}

The Gemini Deep Deep Survey\footnote{For the Public Data Release,
visit http://www.ociw.edu/lcirs/gdds.html}(Abraham et al.\ 2004) is
the deepest survey targeting galaxies in the redshift desert
($0.8<z<2$). It is based on spectra obtained with the Gemini Multi
Object Spectrograph (GMOS), operating in nod \& shuffle mode
(Glazebrook \& Bland-Hawthorn 2001; Cuillandre et al.~1994) for
precise sky subtraction. The survey has been primarily targeting
galaxies with photometric redshifts $z>0.8$ and $K<20.6$, but other
objects were included to fill gaps in the GMOS masks.  The galaxy
sample is complete down to $K$ (Vega) magnitudes $K=20.6$, and $I$
magnitudes $I_{AB}=24.7$. For comparison, the $I_{AB}$ magnitude limit
for the TKRS of KK04 and CFRS of LCS is $I_{AB}\approx24$ and 22.5,
respectively.  For the $z<1$ galaxies, GDDS is much deeper than CFRS
because it was designed to reach $L^\star$ for local galaxies to
$z=2$, the same flux limit is sub-$L^\star$ at $z<1$.  The observed
GDDS $I$ magnitude corresponds to the rest-frame $B$ magnitude for a
galaxy at $z\sim0.7$.

In terms of stellar mass the GDDS is complete (for $z<2$) down to
$M_\star=10^{10.8}$ M$_\odot$ for all galaxies (Glazebrook et
al. 2004). The completeness increases to $M_\star=10^{10.1}$ M$_\odot$
and $10^{9.6}$ M$_\odot$, for star-forming galaxies at $z<2$ and
$z<1$, respectively. More than 300 spectra were taken, of which about
200 are $z>0.3$ galaxies with secure redshifts. The slit aperture used
for the observations was $0.75\times1.1$ arcsec$^2$, and the signal in
the direction perpendicular to the dispersion was extracted over
$\sim0.8$ arcsec. The wavelength range observed is typically 5500--9800
\AA.

For this work, the GDDS sample selection is based on the requirement
that the spectrum of the galaxies (with secure redshift) covers the
spectral interval of the [OII]$\lambda3727$,
[OIII]$\lambda\lambda4959,5007$ and H$\beta$ lines. These are the
lines necessary to determine the metallicity using the $R_{23}$
calibrator (Pagel et al.\ 1979). The redshift range is thus restricted
to $0.40 < z < 0.98$.  Figure~\ref{f1} shows $I_{AB}$ vs.\ $z$ for
GDDS galaxies in this redshift interval (73). Twelve of these are
excluded from the sample because one or more lines are corrupted or
just outside the spectral range (which to some extend depends also on
the position of the slit in the mask); one object is clearly
contaminated by an AGN and another by an interloper, so are excluded
from the sample. Among the 59 remaining objects, the selected 28
(filled squares) show [OII], H$\beta$ and [OIII] emission; 25\% and
75\% are classified as late type and intermediate type, respectively.
The other non-selected galaxies are early type (empty circles) or
intermediate type (empty squares) with only weak or no [OII] emission,
and no other emission line.

Our detection limit for emission line fluxes is a function of redshift
(the spectral sensitivity is higher in the blue than in the red) and
is well represented in the observed redshift interval by the function

\begin{equation}\label{lim}
f_{lim}(10^{-18}\ {\rm erg s^{-1}\ cm^{-2}})= 4.31z-1.12
\end{equation}

\begin{figure*}
\centerline{\epsfxsize=18cm \epsffile{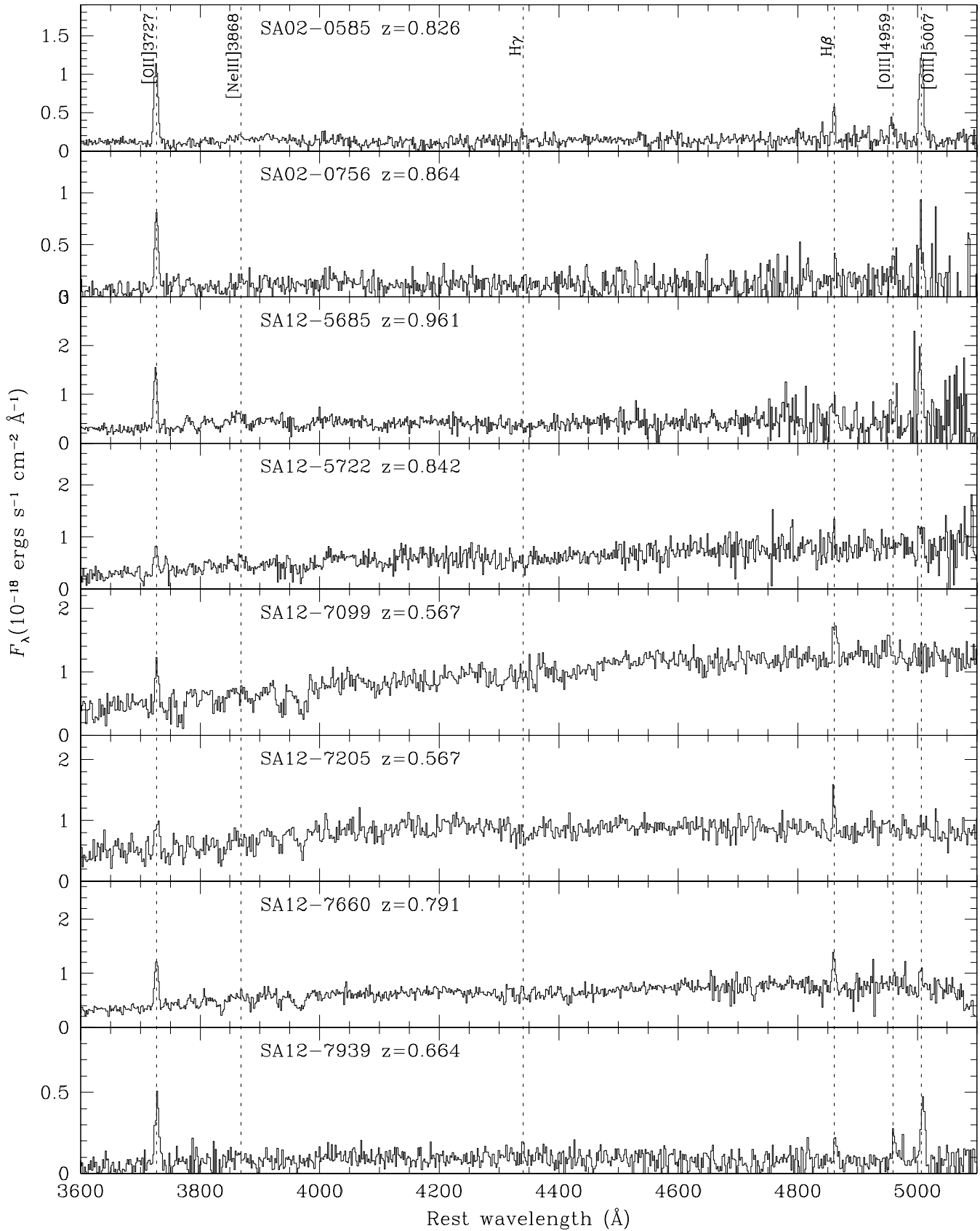}} 
\caption {The sample of
GDDS galaxies at $0.4<z<1$. Relevant emission lines are marked by the
vertical dashed lines.}
\label{f2}
\end{figure*}

\begin{figure*}
\centerline{\epsfxsize=18cm \epsffile{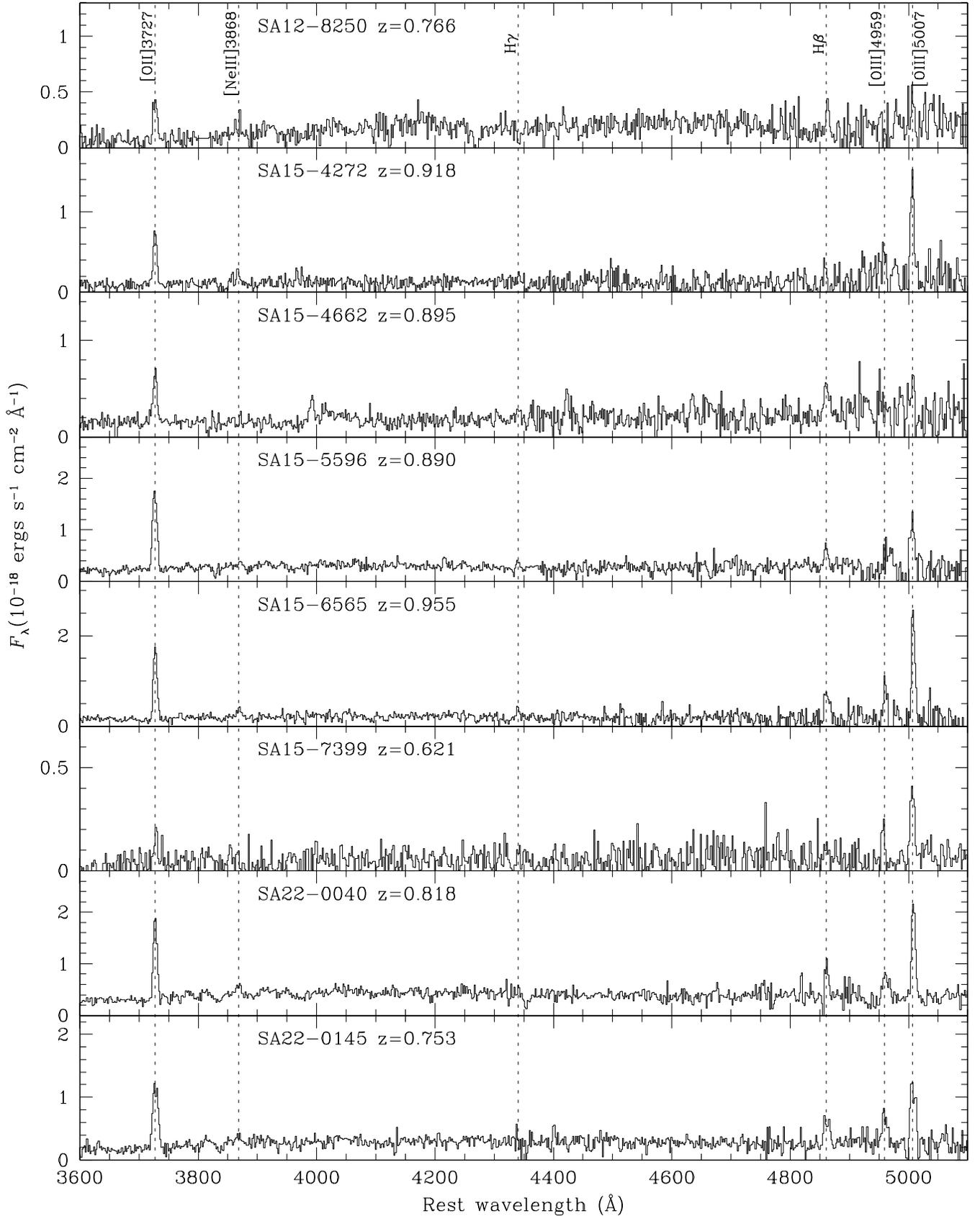}} 
\caption {As in Figure~\ref{f2}.  The spurious emission line at
$\lambda\sim 4000$ \AA\ in the spectrum of SA15-4662 is due to order
overlap in the GMOS mask (see text).}
\label{f3}
\end{figure*}

\begin{figure*}
\centerline{\epsfxsize=18cm \epsffile{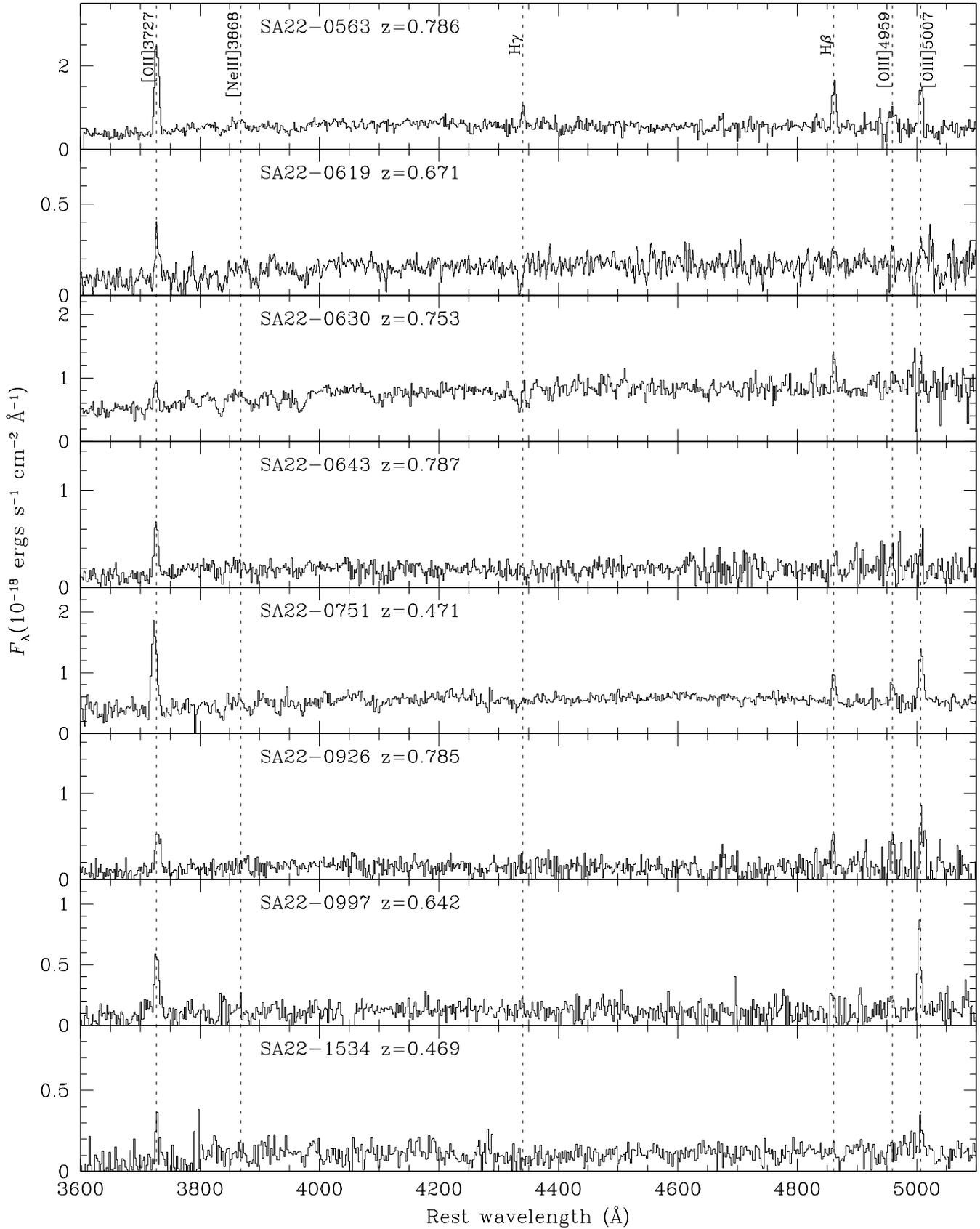}} 
\caption {As in Figure~\ref{f2}.}
\label{f4}
\end{figure*}

\begin{figure*}
\centerline{\epsfxsize=18cm\epsffile{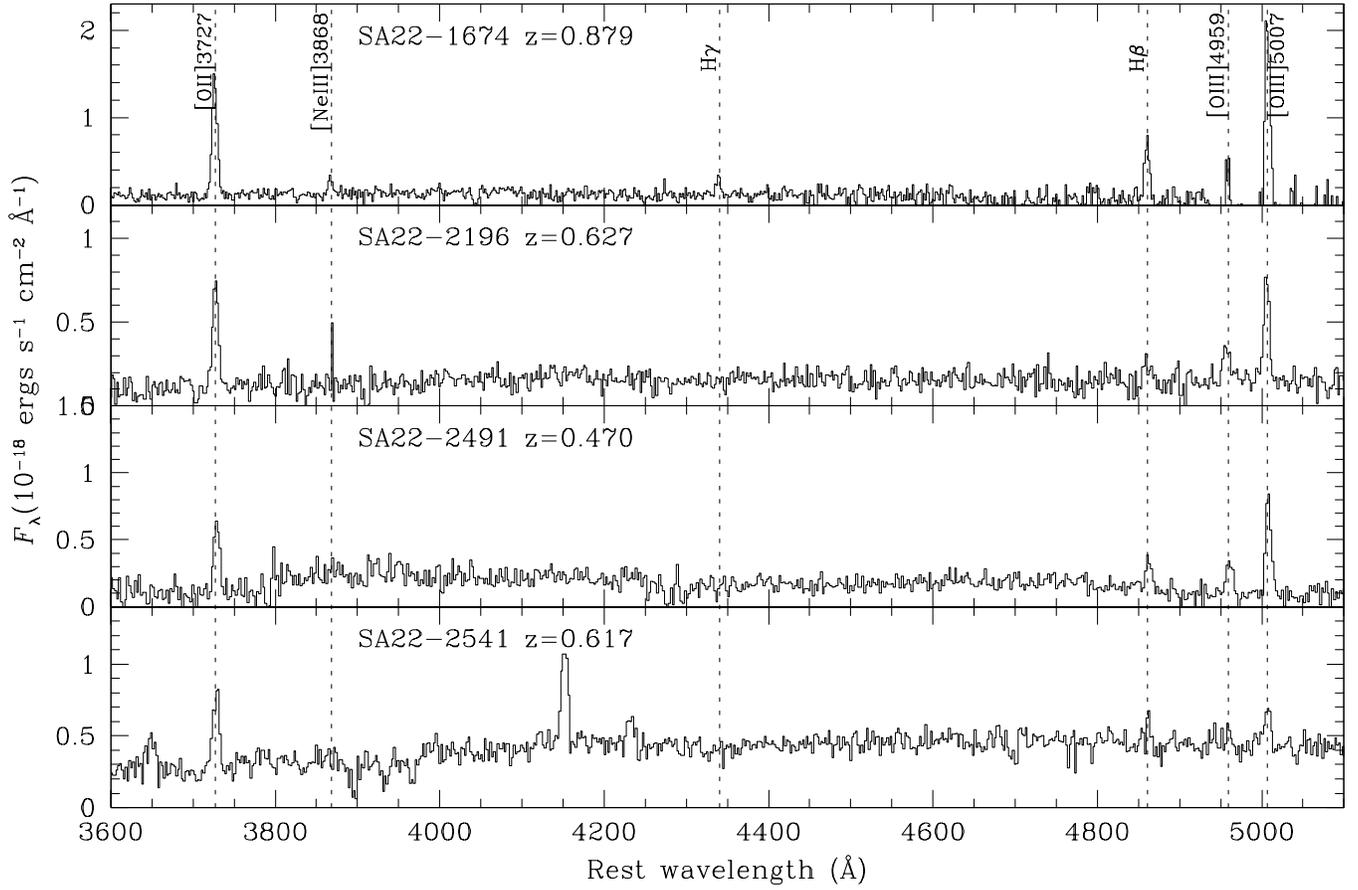}} 
\caption {As in Figure~\ref{f2}. The two spurious emission lines at
$\lambda\sim 4200$ \AA\ in the spectrum of SA22-2541 are due to order
overlap in the GMOS mask (see text).}
\label{f5}
\end{figure*}

\begin{figure*}
\centerline{\epsfxsize=18cm \epsffile{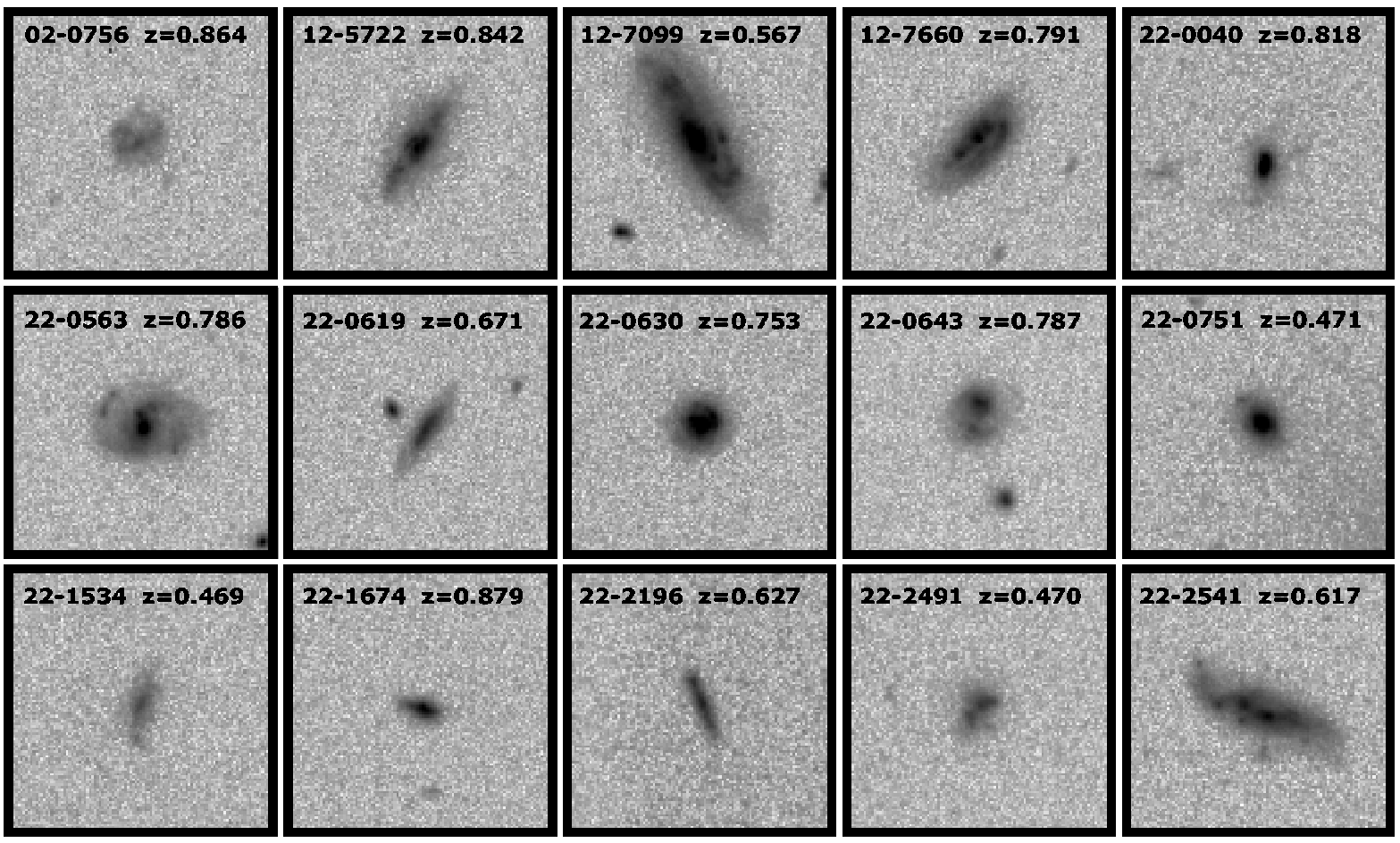}} 
\caption {HST/ACS F814W images of 15 of the 28 GDDS selected
galaxies. The size of the images is $5\times 5$ arcsec$^2$.}
\label{f6}
\end{figure*}

\noindent
We detected line fluxes down to $(0.6-3.2)\times10^{-18}$ erg s$^{-1}$
cm$^{-2}$ in the interval $z=0.4-1$ ($3\sigma$). For 4 galaxies,
the [OIII] or H$\beta$ emission is below detectability, so a
limit was derived for the metallicity.  The CFRS detection limit 
of LCS is $\sim2\times10^{-17}$ erg s$^{-1}$ cm$^{-2}$, i.e.\
a factor of $6-30$ higher than our flux limit.{\footnote{The flux
limit for TKRS is not available, because spectra are not flux
calibrated.}

Spectra of the selected sample are shown in
Figures~\ref{f2}--\ref{f5}. For 15 of the galaxies we also
obtained {\it Hubble Space Telescope} (HST) {\it Advances Camera for
Survey} (ACS) images (Figure~\ref{f6}).  For the majority of them a
disk structure is apparent}. To measure fluxes, the continuum was
estimated from the mean value in two small spectral regions before and
after the line. Errors are derived using the noise spectrum, and
generally agree (within a factor of 1.5) with the pixel-to-pixel
standard deviation in the signal spectra.  For about 1/3 of the
galaxies, the H$\gamma$ line is also detected. All line fluxes, not
corrected for slit aperture losses, are reported in Table~1. The
spectra of SA22-2541 and SA15-4662 show spurious lines due to order
overlap in the GMOS mask (see Abraham et al.\ 2004). Order overlap has
negligible effect on the emission flux, therefore metallicity,
measurements.

The multi-band photometry for the GDDS is provided by the Las Campanas
Infrared Survey (LCIRS; McCarthy et al. 2001; Chen et al. 2002) and is
generally accurate (errors are generally less than 1/10 mag; Abraham
et al.\ 2004).  $K$ and $z'$ band magnitudes (Table~1) are key
parameters used to measure the stellar mass (Glazebrook et al. 2004).
The rest-frame $B$-band absolute magnitude is derived by approximating
the SED between the $V$, $R$ and $I$ bands with a power law.

\section{The GDDS composite spectrum}

We created a composite spectrum from all galaxies in the GDDS sample,
for two reasons. One is to estimate the stellar absorption used to correct
the Balmer emission lines; the other is to estimate an average value of dust
extinction, via the Balmer decrement. Fluxes of Balmer emission lines
in Table~1 are only corrected for the stellar Balmer absorption, and
not for the dust extinction.

\subsection{Stellar Balmer absorption}\label{balmer}

\begin{figure*}
\centerline{\epsfxsize=18.cm \epsffile{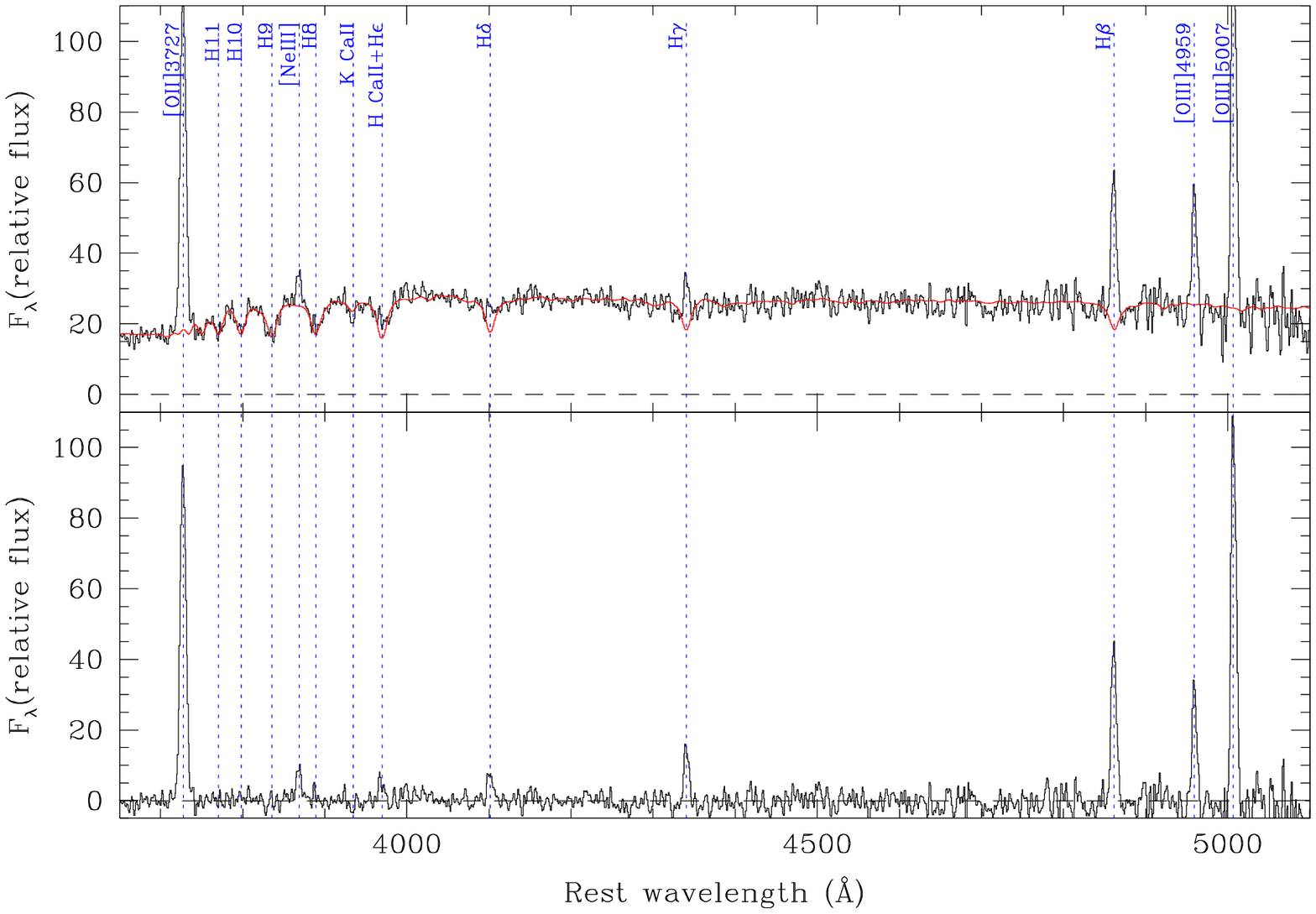}} 
\caption {Composite spectrum of 26 $0.4 < z < 1$ GDDS galaxies. The
dotted vertical lines mark the relevant features. The smooth spectrum
in the upper panel is the ``best-fit'' model (Bruzual \& Charlot 2003)
of the stellar component, obtained for a 50 Myr old stellar population
and a visual extinction $A_V^\star=1.6$ (metallicity is assumed to be
solar). The lower panel shows the emission lines after the stellar
continuum subtraction.  Fluxes measured from the composite spectrum
are listed in Table~2.  The Balmer stellar absorption profile has been
subtracted from each galaxy spectrum before measuring emission line
fluxes, and is of the order of 3.6 and 3.4 \AA\ (equivalent width) for
H$\beta$ and H$\gamma$, respectively, depending on the emission line
FWHM.}
\label{f7}
\end{figure*}

To combine all spectra together, each spectrum has been normalized to
unity, by dividing the flux by its mean value in the interval
$\lambda\lambda=4200-4400$ \AA. Thus each galaxy has the same weight
in the composite spectrum (Figure~\ref{f7}). The best fit to the
stellar continuum, obtained using Bruzual \& Charlot (2003) stellar
population synthesis models, is for a 50 Myr old stellar population
and a visual extinction\footnote{Throughout the paper $A_V^\star$ is
the visual extinction of the stellar continuum, and $A_V$ is the
visual extinction affecting the Balmer emission lines.} for the
stellar continuum $A_V^\star=1.6$. A Calzetti extinction law (Calzetti
2001) and a solar metallicity are assumed. The model gives an
acceptable fit when the dust extinction and the age of the stellar
population span $0.7<A_V^\star<2.0$ and $30-200$ Myr, respectively. If
the dust extinction is in the higher or lower end, only ages below 80
Myr or in the range $100-200$ Myr are acceptable, respectively. In
these intervals, the emission fluxes in the composite generally vary
by less than 5\%.  The best-fit model is only used for the Balmer
absorption correction and does not give a comprehensive description of
the underlying stellar population.  In general, the properties (e.g.,
dust extinction, metallicity and ages) of the stellar component in the
optical (dominated by small and intermediate mass stars) can differ
from those of the young star-forming component represented by the
optical emission lines. Moreover, strictly speaking, the true
$A_V^\star$ is very likely lower than the estimated value, because of
blue-flux spectral loss due to the atmospheric dispersion. This has
little effect on the Balmer absorption features.

From the composite we derive an equivalent width EW correction for
H$\beta$ and H$\gamma$ absorptions of the order of 3.6 \AA\ and 3.4
\AA, respectively, depending on the underlying emission line
FWHM. Changing the H$\beta$ absorption EW by $\pm$\%50 changes the
metallicity by less than 0.1 dex in 83\% of the galaxies, and on
average by 0.06 dex. For the CFRS sample, LCS have used a Balmer
absorption correction of 3 \AA, with 2 \AA\ uncertainty, as
derived for local irregulars and spirals by Kobulnicky, Kennicutt, \&
Pizagno (1999). In extragalactic HII regions the correction is lower,
with a typical EW of 2 \AA\ (McCall, Rybski, \& Shields 1985).

After subtracting the stellar continuum (bottom panel of Figure~\ref{f7}),
emission line fluxes in the composite are measured and reported
(relative to H$\beta$) in Table~2. Errors are estimated by using the
standard deviation in the subtracted spectrum around each emission
line.

\begin{figure}
\centerline{\epsfxsize=9cm \epsffile{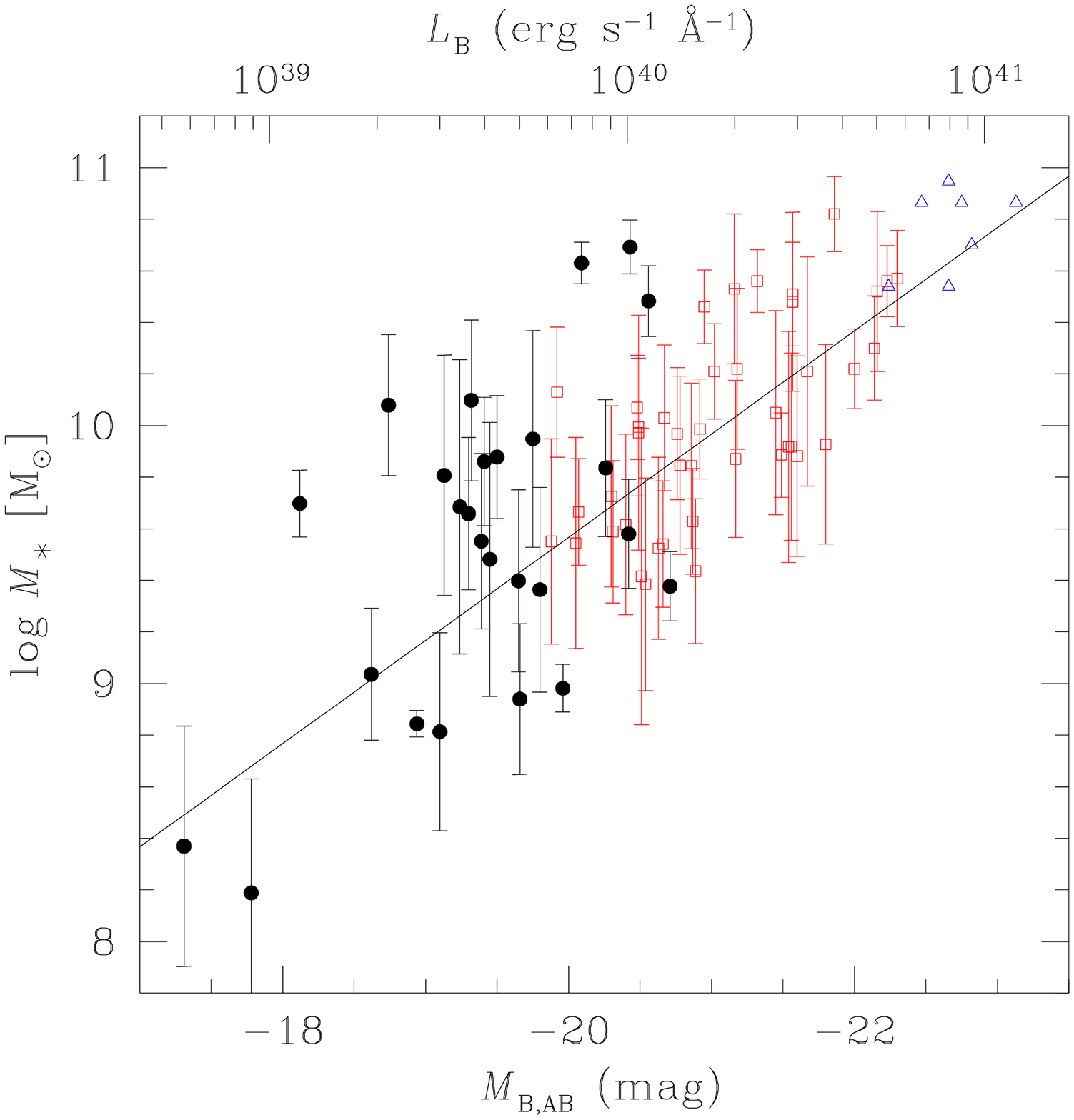}} 
\caption {Stellar Mass
and $B$ absolute magnitude (or luminosity, upper horizontal axis) for
the GDDS (filled circles) and CFRS (empty squares) $z<1$ galaxies. The
open triangles are the $z\sim2.3$ LBGs (Shapley et al.\ 2004) after
correcting the stellar mass for BG03 IMF. The straight line marks a
constant ratio of stellar mass to $B$ luminosity, in solar units, of
$M_{\star,\odot}/L_{B,\odot}=0.3$.}
\label{f8}
\end{figure}

\subsection{Balmer decrement and dust extinction correction}\label{dust}

The comparison between the Balmer decrement (high order Balmer
emission line fluxes relative to the H$\beta$ flux, after stellar
absorption correction) observed in the composite spectrum and the
theoretical value (in the case of no extinction, from atomic physics)
provides a reasonable estimate of the mean dust extinction.  The
theoretical line ratios (Osterbrock, 1989) are only a weak function of
the gas temperature, and are shown in Table~2 for $T=10,000$ $K$ and
5000 $K$.  The visual extinction is estimated assuming the MW
extinction law (practically unchanged for LMC or SMC extinctions). The
error-weighted mean, from H$\gamma$/H$\beta$ through
H8/H$\beta$, is $A_V=2.13\pm0.32$ or $1.92\pm0.32$ for $T=10,000$
$K$ or 5000 $K$, respectively.

In principle the assumed visual extinction can be checked by comparing
line EW ratios (almost unaffected by dust) with the
extinction-corrected flux ratios. This method is not very accurate
because it is based on the assumption that the continuum flux ratio at
the [OII] and H$\beta$ wavelengths is nearly one (Kobulnicky \&
Phillips 2003). Nonetheless, in the subsample of 15 galaxies for which
an accurate estimate of the emission line EWs is possible, we found a
good match between EW ratios and flux ratios for $A_V\sim2$.

We compare our $A_V$ with other estimates derived for different
samples.  Cid Fernandes et al.\ (2005) found an empirical relation
between the visual gas extinction $A_V$ and the visual stellar
extinction $A_V^\star$ using data from the Sloan Digital Sky Survey
(SDSS).  If transformed to take into account a small difference in the
extinction law adopted by us, this relation is:

\begin{equation}\label{av}
A_V = 3.173+1.841A_V^\star-6.418~\log\left(\rm \frac{H\alpha}{H\beta}\right)_{th}
\end{equation}

\noindent
where $(\rm H\alpha/H\beta)_{th}$ is the line flux ratio expected from
atomic physics theory.  Considering that from the composite spectrum
best fit we derived an upper limit for the stellar extinction of
$A_V^\star=1.6$, then $A_V<3.2$ or $<3.0$, for $T=10,000\ K$ or $5000$
$K$, respectively. We also consider that Eq.~\ref{av} is based on data
at $z\sim0.1$ and can vary in the high-$z$ Universe.

An extinction for the ISM similar to ours ($A_V=2.4\pm0.4$) was
derived from the $\rm H\alpha/H\beta$ value (after stellar absorption
correction) measured in the SDSS composite spectrum, the ``cosmic
optical spectrum'' (Glazebrook et al.\ 2003).  LCS assumed $A_V=1$,
and this could be more appropriate for relatively brighter (hence
likely less extincted) galaxies as in the CFRS sample. This gives $\rm
H\gamma/H\beta=0.39$, which is significantly larger than our $\rm
H\gamma/H\beta=0.29\pm0.02$ (Table~2).  An older and more extincted
stellar population model (older than 100 Myr, and $A_V^\star>1.5$)
would give a higher $\rm H\gamma/H\beta$ in our composite, then a
lower $A_V$. However, the fit is much worse, the $\chi^2$ is 3 times
higher.

Calzetti (1997) found a mean visual extinction in a sample of 19
starburst galaxies of $A_V=1.35$ with a 1$\sigma$ dispersion of 0.77
magnitudes. The higher value in our sample can be an indication of a
higher dust content in galaxies that are generally more massive and
more metal rich than the local starbursts (Brinchmann et al.\ 2004).

The H$\gamma$ emission line is detected in 10 GDDS galaxies (Table~1),
so the Balmer decrement provides the dust extinction in these spectra
individually. For this subsample we found $<A_V>=1.66$ with a
dispersion of 1.3 magnitudes, indicating that galaxies for which
H$\gamma$ is detected tend to be less extincted. However, because the
errors are always higher than 0.5 mag, we ignore $A_V$ estimated this
way. It is worth noticing that the median stellar mass for these
galaxies ($M_\star=10^{9.4}$ M$_\odot$) is lower, but close, to the
median value for the whole sample ($M_\star=10^{9.6}$ M$_\odot$),
suggesting that if the dust extinction were larger for more massive
galaxies, this effect is not apparent in our intermediate redshift
sample.

The visual extinction uncertainty (0.3 mag) does not affect very much
the metallicity estimates from line ratios because the wavelength
baseline of the emission lines used (from [OII] to [OIII]) is not
very large ($3727-5007$ \AA). Consequentially our results are
relatively unaffected for small variations of $A_V$. We will adopt
$A_V=2.1\pm0.3$ and discuss results for $A_V=1$.  In the GDDS sample the
metallicity decreases on average by 0.2 dex if $A_V$ goes from 0 to 3.

\begin{figure}
\centerline{\epsfxsize=9cm \epsffile{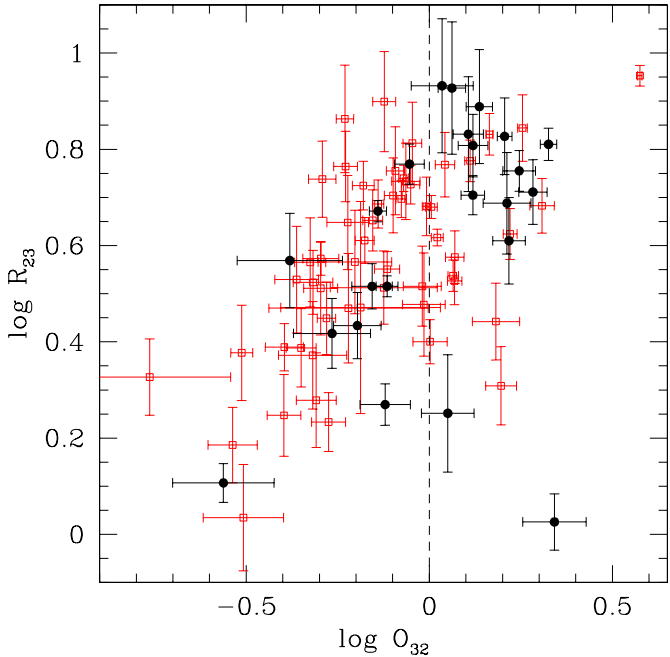}} 
\caption {$R_{23}$
vs. $O_{32}$ for the GDDS ({\it filled circles}) and the CFRS
({\it open squares}) galaxies.  The dust correction is not
applied.  The CFRS points are calculated using emission line fluxes
from LCS.  The dashed line marks the region with $O_{32}=1$. The large
fraction of galaxies with $O_{32}>1$ in GDDS with respect to CFRS is
an indication of lower metallicity, and/or higher dust extinction,
and/or higher ionization. The $O_{32}$ value decreases by
0.4 dex if the dust correction goes from $A_V=0$ to 3.}
\label{f9}
\end{figure}

\section{Stellar masses}\label{mass}

The stellar mass of galaxies in the GDDS and CFRS samples have been
estimated using the procedure described in Glazebrook et al.\ (2004)
and also used in Juneau et al.\ (2005). The galaxy SED is modeled
using the multi-band photometry.  Galaxy model spectra are provided by
P\'EGASE.2 (Fioc \& Rocca-Volmerange 1997, 1999). The fitted
parameters are the stellar dust extinction (in the range $0<A_V<2$),
the age of the stellar population, and the metallicity. The star
formation rate is assumed to decline exponentially, with e-folding
time ranging from $\tau=0.1$ to 500 Gyr, combined with a bursty
component with $\tau=100$ Myr and variable mass fraction. The best-fit
SED gives a $M/L_K$, from which masses are reliably derived. The IMF
used is that derived by Baldry \& Glazebrook (2003; hereafter BG03),
and gives a total stellar mass 1.8 or 1.2 times smaller than for a
Salpeter or a Kroupa IMF, respectively. The difference with Salpeter
IMF is mainly in the low mass stars. The error in the stellar masses
includes uncertainties on $A_V$, $\tau$ and mass fraction of the
burst. Our mass estimates, based on the rest-frame optical-NIR
photometry, is much more robust than when based on rest-frame UV
light, which is strongly affected by very uncertain parameters like
dust extinction and recent star formation.  This SED-fitting method
gives a mass accuracy generally better than a factor of two for the
$z>0.8$ GDDS galaxies (Glazebrook et al.\ 2004).  A similar method was
recently tested against other stellar mass estimates, and was found to
be generally robust and accurate (Drory, Bender \& Hopp 2005).

For the GDDS galaxies in our $z<1$ sample, the $VIz'K$ photometry
allowed reliable mass estimates to be derived for 27 galaxies, out of
28 (Table~1).  For one galaxy, the stellar mass could not be
calculated for incomplete photometric information. The median error
is 0.3 dex, and the errors are lower than a factor of two and three
for 55\% and 94\% of the objects, respectively.  For many galaxies,
only upper limits to the $K$ magnitude are available, therefore stellar
masses rely on the $z'$ magnitude. These masses were not derived in
Glazebrook et al.\ (2004), but their uncertainties are still
acceptable, as the observed $z'$ band corresponds to the rest-frame
$V$ band at redshift $\sim0.7$.  These errors are typically $\sim0.15$
dex higher than those for galaxies with measured $K$ magnitude. In three
cases for which neither $K$ nor $z'$ are available, mass errors are
larger.

The sample median redshift and redshift interval are $z=0.79$ and
$0.47<z<0.96$, respectively.  The lowest and the highest masses are
$M_\star=10^{8.2}$ M$_\odot$ and $10^{10.7}$ M$_\odot$, and the sample
is complete down to $M_\star=10^{9.6}$ M$_\odot$.

For the CFRS sample of 69 galaxies, $VIK$ photometry is used and
masses are measured for 42 galaxies  (Table~3). The mass error is
always smaller than 0.6 dex (less than a factor of 4), and the median
error is 0.3 dex.  The lowest measured mass is $M_\star=10^{9.4}$
M$_\odot$, i.e.\ 16 times higher than the lowest stellar mass in the GDDS.
The highest mass is $10^{10.8}$ M$_\odot$. The median redshift and
interval are $z=0.68$ and $0.479<z<0.915$, respectively.

The total number of galaxies for which we measured masses is 69. For
56 of these we also determined metallicities (see \S\ref{Z}).
 
Figure~\ref{f8} shows the stellar mass and the $B$ absolute magnitude
(or luminosity) for the GDDS and CFRS samples. It is apparent that
GDDS galaxies are less massive (on average 2.24 times) and fainter
($\sim1.75$ magnitudes, or a factor of 5 in flux) than CFRS galaxies.
GDDS galaxies are on average $>1.1$ mag fainter in $K$ than CFRS
galaxies. The fact that the ratio between the mean CFRS $B$ luminosity
and the mean GDDS $B$ luminosity is larger than the mass ratio, 
indicates that GDDS galaxies are
on average more dust extincted than CFRS galaxies.

As a reference, the LBGs at $z\sim2.3$ (Shapley et al.\ 2004, also in
Figure~\ref{f8}) are much brighter and more massive than the CFRS
galaxies. (LBG masses are scaled to take into account the different
IMF used, and the recycling into the ISM, which Shapley et al.\
did not include and which makes the final stellar masses 1.4 times smaller.)

\begin{figure}
\centerline{\epsfxsize=9cm \epsffile{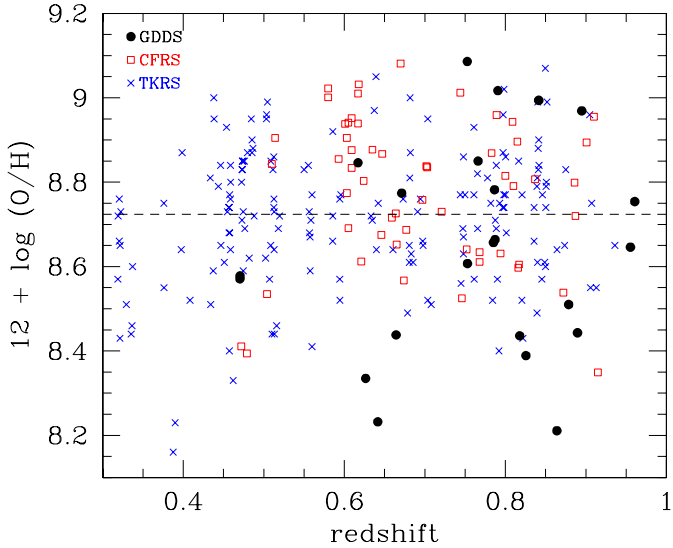}}
\caption{Metallicity as a function of redshift for GDDS (filled
circles), CFRS (empty squares) and TKRS (crosses, from KK04).  The
mean metallicities in the GDDS, CFRS and TKRS are $12+\log(\rm
O/H)=8.63\pm0.26$, $8.78\pm0.17$ and $8.72\pm0.17$, respectively. The
mean total metallicity (dashed line) is $12+\log(\rm
O/H)=8.72\pm0.17$, i.e. roughly solar.  We do not detect any
significant redshift evolution for the total sample of 261 galaxies,
from $z=0.32$ to $z=0.96$ (time interval of 4.3 Gyr).}
\label{f10}
\end{figure}

\section{The metallicity using the R$_{23}$ parameter}\label{Z}

Emission line fluxes of [OII]$\lambda3727$, H$\beta$, and
[OIII]$\lambda\lambda4959,5007$ provide an estimate of
metallicity\footnote{We express metallicity in terms of the number of
oxygen atoms relative to hydrogen: $12+\log (\rm O/H)$.}  through the
$R_{23}$ calibrator. When the [OIII]$\lambda4959$ emission line is
barely detected, the flux is assumed from atomic physics to be 0.34
times the [OIII]$\lambda5007$ flux, and the error propagated
accordingly.  The $R_{23}$ parameter, as defined by Pagel et al.\
(1979), is

\begin{equation}\label{r23}
\log R_{23} = \log \left(\frac{f_{\rm [OII]\lambda3727}+f_{\rm [OIII]\lambda\lambda4959,5007}}{f_{\rm H\beta}}\right) \equiv x
\end{equation}

\noindent
We  also define the $O_{32}$ parameter: 
\begin{equation}\label{o32}
\log O_{32} = \log \left(\frac{f_{\rm [OIII]\lambda\lambda4959,5007}}{f_{\rm [OII]\lambda3727}}\right)\equiv y
\end{equation}

\noindent 
used to correct for the ionization. The $R_{23}$ and $O_{32}$ for GDDS
galaxies are reported in Table~1 and are corrected for dust
extinction. When dust extinction is applied, $R_{23}$ increases and
$O_{32}$ decreases. $O_{32}$ decreases by 0.4 dex if one corrects for
$A_V$ in the range $0-3$.

The ionization-sensitive $O_{32}$ not corrected for dust in GDDS and
CFRS (Figure~\ref{f9}) is $>1$ in 62\% \& 27\% of the galaxies,
respectively.  In the $\sim200$ TKRS galaxies, KK04 found that 32\%
have $O_{32}>1$. However, because KK04 used equivalent widths (not
very sensitive to dust extinction) this fraction would be higher if
unextincted fluxes are used instead.  As noted by Kewley \& Dopita
(2002), $O_{32}$ is higher not only in highly ionized or dusty
regions, but also in low metallicity regions.  In fact, in high
metallicity regions, most of the nebular cooling occurs through far-IR
oxygen lines, so the [OIII] emission becomes weak and the $O{32}$
parameter decreases.

The generally higher $O_{32}$ in the GDDS with respect to CFRS
(Figure~\ref{f9}) suggests that GDDS galaxies are more ionized and/or
more extincted and/or less metal rich than CFRS galaxies. Additional
evidence for more extinction in GDDS was provided in \S~\ref{mass}.

The LBGs studied by Pettini et al.\ (2001) all have $O_{32}>1$ and
$R_{23}> 5$. The difference between these and our $z\sim0.7$ galaxies
likely implies lower metallicities for LBGs, not surprising at higher
redshifts.

\subsection{The GDDS and CFRS metallicity}

The average metallicity of galaxies can be derived through different
calibrators which use $R_{23}$ and $O_{32}$.  Initially proposed
by Pagel et al.\ (1979), the $R_{23}$-metallicity relation was
reformulated by several authors with the aid of the
ionization-sensitive parameter $O_{32}$ (Edmunds \& Pagel 1984; Dopita
\& Evans 1986; McGaugh 1991; Pilyugin 2001; Kewley \& Dopita 2002;
KK04). The difference between the different calibrators can be as high
as 0.2 dex, with the theoretical methods giving generally higher
metallicities than the empirical methods (KK04).  Another limit of the
$R_{23}$ calibrator is its double solution. In fact, at low
metallicity, when the electron temperature $T_e$ is high enough to
keep the gas collisionally ionized, $R_{23}$ scales with
metallicity. But when the metallicity reaches a limit, the gas cooling
occurs through the far-IR oxygen lines, then the $R_{23}$-metallicity
dependence turns around, i.e.\ $R_{23}$ decreases for higher
metallicity.  This happens when $R_{23}$ is of the order of 10 and
$12+\log (\rm O/H) \sim 8.3$. In the turnover region, the metallicity
is easily $0.2-0.4$ dex more uncertain because the two solutions are
close and both possible. To break the degeneracy, other methods
requiring other lines can be used, for instance [NII] and [SIII]
(Denicol\'o, Terlevich, \& Terlevich 2002; Pettini \& Pagel 2004;
Bresolin et al.\ 2005). However, at high redshift this requires
expensive NIR spectroscopy, not easily accessible. Because $T_e$ is
sensitive to metallicity, the [OIII]$\lambda4363$/[OIII]$\lambda5007$
temperature-sensitive ratio gives the most reliable method (Kobulnicky
et al.\ 1999).  However [OIII]$\lambda4363$ is a weak line, useful
only when the spectral S/N is high.

When the only available set of lines is [OII], [OIII] and H$\beta$,
as in our and all cases where optical spectroscopy is available  for high-$z$
galaxies, the mass and the age of the galaxy can be used to break the
degeneracy, as massive and evolved galaxies are typically metal rich.  As
the galaxies in our sample are generally massive, and our $R_{23}$ is
generally far from the turnover region, we will only consider the
upper branch solution. This is supported by results on galaxies with
similar luminosities and redshifts in TKRS and CFRS, for which other
metallicity diagnostics indicated that the upper branch solution was
correct (KK04, LCS).

Among the different calibrators, the empirical one proposed by
McGaugh (1991) is the most commonly adopted. KK04 compared
some of them, discussed possible biases, and concluded that the best
one is represented by a mean value between McGaugh's and the 
theoretical one given by Kewley \& Dopita (2002). The new calibrator
is approximated, for metallicities higher than $12+\log (\rm
O/H)=8.4$, by the following expression:

\begin{eqnarray}\label{met}
12+\log({\rm O/H}) = 9.11-0.218x-0.0587x^2-0.330x^3-\nonumber \\
+0.199x^4-y(0.00235-0.1105x-0.051x^2-0.04085x^3-\nonumber \\
+0.003585x^4)\nonumber \\
\end{eqnarray}

\noindent
Roughly speaking, the metallicity of Eq.~\ref{met} is not too
different from McGaugh's, on average 0.07 dex higher in the range $8.4
< 12+\log(\rm O/H) <9.3$.

For this work, we used Eq.\ \ref{met}. For 4 galaxies, $R_{23}$
is in the turnover region, and the metallicity is more
uncertain. In these cases (when $12+\log (\rm O/H)< 8.4$) we used the
McGaugh calibrator and applied the small positive correction of $0.07$
dex (as derived from the comparison with KK04).  These galaxies do not
change much our findings on the M-Z distribution at $z\sim0.7$ (see
\S\ref{mz}).  GDDS metallicities are reported in Table~1 (calculated
for $A_V=2.1\pm0.3$). The mean value (and $1\sigma$ dispersion) is
$12+\log (\rm O/H)=8.63\pm0.25$.  This is only about 0.07 dex higher
for $A_V=1$.  Errors on metallicities include the emission flux
and $A_V$ uncertainties only. The mean error, is 0.12 dex. Systematic
uncertainties are generally higher than the measured errors for large
metallicities.  For 4 galaxies with an upper limit on H$\beta$ or
[OIII] flux, a lower or upper limit on the metallicity are provided,
respectively.

LCS used for the CFRS sample the calibrator provided by McGaugh (1991)
and $A_V=1$.  For consistency with GDDS, we have re-calculated
metallicities, applying Eq.~\ref{met} (Table~3).  Errors in the
metallicities include an arbitrary uncertainty on $A_V$ of 0.3 mag.  Also
for this sample, galaxies with an upper limit on H$\beta$ or [OIII]
fluxes provide a lower or upper limit on the metallicity,
respectively. Two galaxies with no line detection, have no information
on the metallicity.  Table~3 also lists old LCS metallicities for the
upper branch. Our new mean metallicity and error are $12+\log (\rm
O/H)=8.78\pm 0.17$ and 0.07 dex, respectively (the LCS mean is 0.05
dex lower).

Figure~\ref{f10} shows $12+\log(\rm O/H)$ as a function of redshift,
for the GDDS, CFRS and TKRS samples. The mean metallicity of all 261
galaxies is $12+\log(\rm O/H)=8.72\pm0.17$, i.e.\ roughly
solar\footnote{We assumed as a solar metallicity the value provided by
Allende Prieto et al.\ (2001), which is $12+\log(\rm
O/H)_\odot=8.69$.}. Although the redshift range ($0.3<z<1.0$)
corresponds to a time interval of 4.3 Gyr, no redshift evolution is
detected (i.e.\ no significant metallicity-redshift correlation). We
emphasize that galaxies in this plot include all masses, therefore the
presence of a mass-metallicity relation at these redshifts (see
\S\ref{mz}) may obscure any redshift evolution of metallicity.

\begin{figure}
\centerline{\epsfxsize=9cm \epsffile{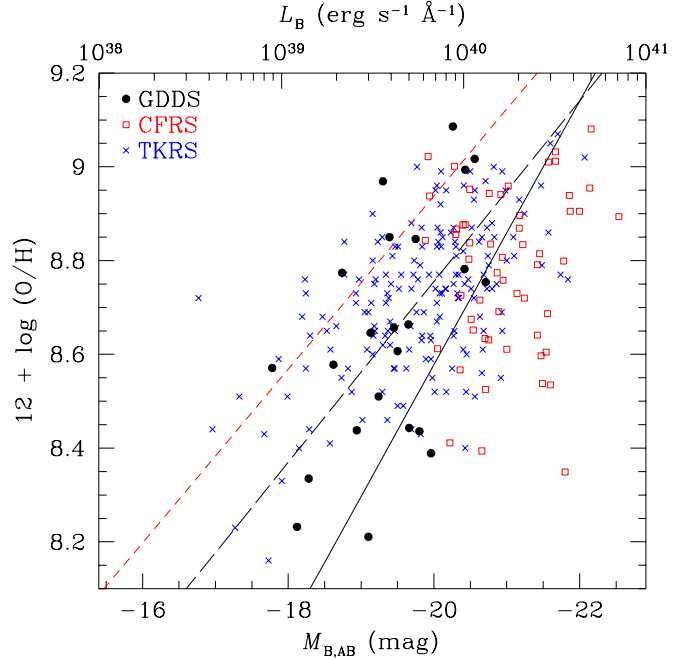}} 
\caption {$12+\log (\rm O/H)$ as a function of $B$ absolute magnitude
$M_{B,AB}$ (or luminosity, upper horizontal axis) for the GDDS (filled
circles), CFRS (open squares) and TKRS (crosses; KK04). The linear
bisector fits (see text) for GDDS+CFRS and TKRS are shown as solid and
long-dashed lines, respectively.  The short-dashed line is the
luminosity-metallicity relation found for the SDSS sample at
$z\sim0.1$ (Tremonti et al. 2004). }
\label{f11}
\end{figure}

\begin{figure}
\centerline{\epsfxsize=8.5cm \epsffile{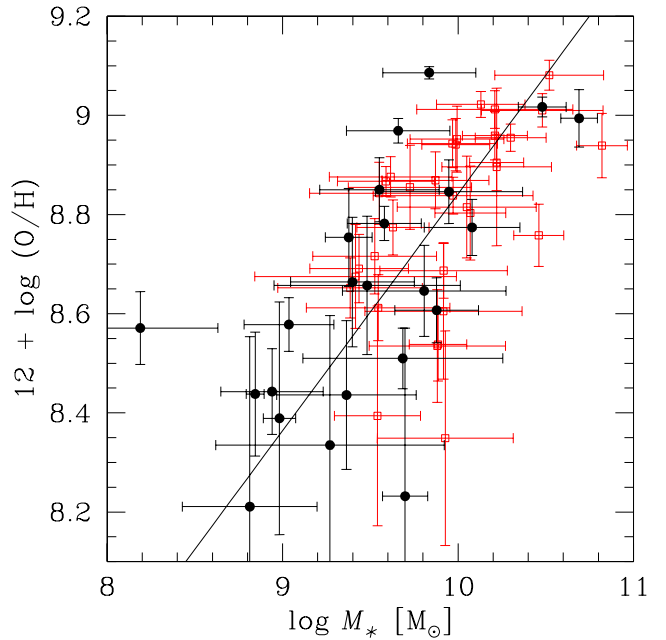}}
\caption{Metallicity as a function of stellar mass for GDDS (filled
circles) and CFRS galaxies (open squares). Error bars include
measurement errors but not possible systematic effects. The straight
line is the linear bisector fit. The correlation is more than
6$\sigma$ significant. The metallicity scale is as in Figure~\ref{f11},
and the mass scale spans three orders of magnitudes, as in the
luminosity scale of the upper $x$-axis of Figure~\ref{f11}. This way it
is possible to visually compare the point scatter in the two figures.}
\label{f12}
\end{figure}

\section{Luminosity-metallicity relation at $z\sim0.7$}

In Figure~\ref{f11} we show metallicity vs.\ $B$ absolute magnitude (or $B$
luminosity) for the GDDS (filled circles), CFRS (open squares) and the
TKRS (crosses; KK04) galaxies.  The GDDS+CFRS sample (79 galaxies)
shows a 4$\sigma$ significant correlation. The linear bisector best
fit (Feigelson \& Babu 1992) is expressed by:

\begin{equation}
12+\log ({\rm O/H}) = (-0.280\pm0.045) M_{\rm B,AB}+(2.977\pm0.937)
\end{equation}

\noindent
The fit does not change much (although the significance is slightly
lower) if $A_V=1$ is assumed for the GDDS galaxies.

In Figure~\ref{f11}, we show this bisector fit (solid line) together
with the one found by KK04 for  177 TKRS galaxies [$12+\log ({\rm
O/H}) = -0.193 M_{\rm B,AB}+4.900$, long-dashed line]. The short-dashed
line is the L-Z relation for the SDSS 53,000 
$z\sim0.1$ galaxies (Tremonti et al.\ 2004; hereafter T04),
which is very similar to the one found by Jansen et al.\ (2000) for
the local Nearby Field Galaxy Survey (NFGS).

KK04 note that galaxies in their sample have higher luminosities with
respect to local galaxies with similar metallicities, suggesting a
redshift evolution of the L-Z relation.  This is confirmed by the
GDDS+CFRS sample, but our low-luminosity galaxies seem to have lower
metallicities than the TKRS galaxies.  This can only in part be due to
the slightly different mean redshifts (0.71 for GDDS+CFRS and 0.62 for
TKRS) because the difference corresponds to only 0.5 Gyr. KK04 observe
a steepening of the slope of the bisector fit from $-0.16$ to $-0.24$
(but this include the CFRS sample) going from $z\sim0.3$ to
$z\sim0.9$.  The difference might be related to the dust extinction
correction, which can be a function of the stellar mass. While for the
TKRS galaxies, EWs have been used which are almost insensitive to the
dust correction, for our emission fluxes we have assumed a constant
extinction correction both for GDDS and CFRS. If instead the dust
correction is higher for more massive (than generally brighter)
galaxies, the slope of the bisector fit would be lower. We also note
that the large uncertainties and the different selection criteria for
the three samples can explain the discrepancy.

\section{Mass-metallicity relation at $z\sim0.7$}\label{mz}

Stellar masses and metallicities have been derived for a total of 57
GDDS and CFRS galaxies. One galaxy, SA15-4272, was removed from
the GDDS sample for the large uncertainty on the H$\beta$ flux.  The
median and mean redshift of the remaining 56 galaxies is $z=0.69$ and
0.71, for GDDS and CFRS respectively. The stellar mass and metallicity
distribution (Figure~\ref{f12}) are strongly correlated, at the level
of $6.3\sigma$.  The linear bisector fit gives:

\begin{equation}\label{met2}
12+\log ({\rm O/H}) = (0.478\pm0.058) M_\star+(4.062\pm0.579)
\end{equation}

\noindent
The significance of the correlation is almost the same ($6.6\sigma$)
if the 4 galaxies in the turnover region ($12+\log (O/H)<8.4$) are
removed from the sample. The bisector slope becomes shallower, but
consistent, with the slope for the whole sample ($0.409\pm0.050$).

The scatter around the fit is $\sim0.2$ dex. The scatter for the
luminosity-metallicity distribution of Figure~\ref{f11} is 65\% higher.
Results do not change much if $A_V=1$ for the GDDS galaxies.  The
low-redshift half of the sample ($z<0.7$) is not significantly
differently distributed than the high-redshift half of the sample
($z>0.7$).

We compare this distribution with the M-Z relation recently found by
T04 for the $z=0.1$ SDSS galaxies.  To do that, we need to convert
their relation to one consistent with our choice of IMF and
metallicity calibrator.

T04 used the IMF provided by Kroupa (2001). The conversion from a
Kroupa IMF to a BG03 IMF gives $M_{\star, \rm BG03}=
M_{\star,\rm K01}/1.2$, or a difference of 0.08 dex.

T04 measured metallicities by modeling the emission lines and stellar
continuum, using the approach described by Charlot et al. (in
preparation), while for our sample we used the KK04 calibrator. A
comparison between the two methods indicates that the T04
metallicities are at the most $\sim0.1$ dex higher (this is mainly for
the massive galaxies).  The M-Z polynomial
fit for the SDSS galaxies was recalculated using the KK04 calibrator
(C. Tremonti, private communication).

The ``converted'' T04 M-Z relation (where the KK04 metallicity and
BG03 IMF are used) is:

\begin{equation}\label{newT}
12+\log ({\rm O/H}) = -2.4412+2.1026 \log M_\star -0.09649 \log^2 M_\star
\end{equation}

\begin{figure}
\centerline{\epsfxsize=8.5cm \epsffile{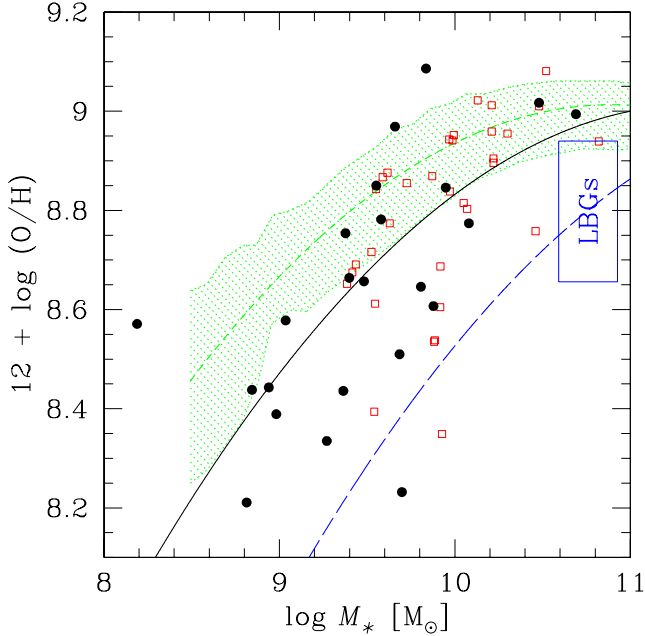}}
\caption {Metallicity as a function
of stellar mass for GDDS (filled circles) and CFRS (open squares). The
points are as in Figure~\ref{f12}, but error bars are omitted.  The
green short-dashed line and hatched area indicate the polynomial fit and
$\pm1\sigma$ dispersion, respectively, derived for the SDSS $ z\sim0.1$
galaxies (T04).  The blue rectangle is the region occupied by the LBGs
at $2.1<z<2.4$ (Shapley et al. 2004) after correcting from the N2 to the
$R_{23}$ calibrator. The black-solid and blue
long-dashed lines are the SDSS polynomial shifted to the right
to match the galaxy distributions at $z\sim0.7$ and $z\sim2.3$,
respectively.}
\label{f13}
\end{figure}

\noindent
The overall difference with the T04 polynomial is not very large.  The
new relation and the $\pm1\sigma$ dispersion are shown in
Figure~\ref{f13}.

In the same figure, we show our $z\sim0.7$ galaxies (as in
Figure~\ref{f12}, but error bars are omitted).  The majority of the
galaxies are distributed below the $z\sim0.1$ relation, and
preferentially these are galaxies with $M_\star<10^{10}$ M$_\odot$, 
suggesting that massive galaxies reach high metallicities much
earlier than low massive galaxies. In other words, the
mass-metallicity relation evolves from being steep at high redshifts
to being flatter at low redshifts.  This picture of massive galaxies
being more quiescent in recent times than less massive galaxies is
also favored by recent results on the SFR density of the Universe
calculated or measured for different galaxy stellar masses (Heavens et
al.\ 2004; Juneau et al.\ 2005) according to which massive galaxies
ceased forming stars (hence stopped enriching the ISM with metals)
much earlier than low-mass galaxies. Low-mass galaxies have a
long-lasting star-formation activity, and significantly produce stars
and metals even in recent times. Those galaxies in the
mass-metallicity diagram of Figure~\ref{f13} will migrate with time
from the left-bottom region to the middle region. A simple closed-box
model which can explain this, is discussed in \S\ref{cb}.

Our conclusion is not inconsistent with the relatively higher
metallicities found for the generally massive LBGs (Shapley et al.\
2004) and the star-forming galaxies at $z\sim2$ of the K20 (de Mello
et al.\ 2004).  The rectangle in Figure~\ref{f13} represents the
region occupied by the LBGs. Masses have been converted to match our
IMF and include recycling into the ISM (if recycling is ignored,
stellar masses are 1.4 times higher). Metallicities are also converted
to match our metallicity calibrator. Shapley et al.\ (2004) used the
N2 calibrator (Pettini \& Pagel 2004), which gives, for those stellar
masses, metallicities $\sim1.4$ to 2.3 times smaller, from the less
massive to the most massive galaxy, than the KK04 calibrator
(C. Tremonti, private communication). On average the difference is
$\Delta [12+\log(\rm O/H)] = -0.23$.

\section{Possible systematic effects}\label{systematics}

One of the puzzling results of the M-Z relation obtained for the
star-forming galaxies by T04 and here, is that these galaxies appear
to be generally metal rich for their stellar mass.  For instance, the
stellar mass of the Milky Way is $\sim5\times10^{10}$ M$_\odot$
(Portinari et al.\ 2004). If its metallicity is approximately solar,
$12+\log (\rm O/H) \sim 8.7$, this is 0.4 dex lower than that
predicted by the M-Z relation of the SDSS ($12+\log (\rm O/H) \sim
9.1$). 

First we have to consider that there are systematic differences
between different metallicity calibrators (see \S\ref{metcal}).
Moreover, it is well known that the metallicity decreases with the
distance from the galaxy center (Garnett et al.\ 1997).  According to
Rolleston et al.\ (2000), in the Milky Way the oxygen
abundance\footnote{This is derived using absorption lines in the
photosphere of B-type dwarf stars.}  goes from $12+\log (\rm O/H)=9$
to 8.6, at 6 kpc and 12 kpc from the center, respectively, with a
radial gradient of $-0.067$ dex kpc$^{-1}$. Although the
quantitative result depends on the empirical method used to measure
metallicity, and although some extragalactic studies indicate lower
central metallicities (Kennicutt et al.\ 2003), it is likely that the
metallicity is higher than the average if the slit aperture used to
collect the galaxy signal is small.  The fiber aperture for the SDSS
sample ($3\arcsec$) corresponds to a median diameter of 4.6 kpc
(T04). At a distance of 4.6/2 kpc from the center, according to
Rolleston et al.\ (2000), the abundance of the Milky Way is 9.2, and
higher if one takes the mean abundances in the inner region. 
Thus it is not surprising that SDSS galaxies have high metallicities.

In the following subsections we show that other possible systematic
effects are not going to dramatically change our conclusions on
the evolution of the M-Z relation.

\subsection{Aperture effects}

If there is a gradient of the metallicity with the distance from the
center (Vila-Costas \& Edmunds 1992; Zaritsky, Kennicutt \& Huchra
1994; Garnett et al.\ 1997; Rolleston et al.\ 2000), it is important
to make sure that this effect is not dominating when comparing samples
obtained using different slit apertures.

We considered that the central part of a galaxy is more luminous than
the outer part, and at high redshift is going to dominate the signal
in the spectra.  Kewley et al.\ (2005) have investigated the aperture
effects as a function of redshifts, and found that the metallicity is
generally not more than 0.1 dex larger than the average in the $\sim4$
kpc central region. This 4 kpc corresponds to an angular size $<0.8''$
for galaxies at $z>0.4$. For this reason, the aperture effects are not
strongly affecting the $z\sim0.7$ results, because the region probed
is generally large enough, and because residual offsets are affecting
galaxies in a similar way.

The SDSS fiber aperture corresponds to a median projected 
diameter of 4.6 kpc (circular area is 16.6 kpc$^2$), with a range
extending from 1 to 12 kpc$^2$, from the low to the high redshift
end. However, T04 estimated a modest (0.1 dex) metallicity change in
the same redshift interval. The GDDS extraction aperture rectangle is
$0.75\times 0.8$ arcsec$^2$ and corresponds to $3.9\times4.2$ kpc$^2$
and $5.9\times6.3$ kpc$^2$, from $z=0.38$ to 0.96. The median for the
sample is $5.6\times5.9$ kpc$^2$. The dispersion of the physical area
in the SDSS sample is higher than in GDDS, and the median value is
lower. It is hard to estimate precisely the mean difference
between the two apertures due to the large aperture dispersion in the
SDSS, we do not expect the aperture effect to be important.

For the CFRS sample, the slit aperture is twice the GDDS aperture in
the dispersion direction (1.3 arcsec, or 9.3 kpc at $z=0.7$) and
20$\arcsec$ in the perpendicular direction, so the galaxy area probed
is larger than the SDSS area. However, following Kewley et al.\
(2005) and considering the small metallcity dispersion observed in the
SDSS sample, we expect the effect to be small.

Another related effect is the galaxy inclination. T04 estimated 0.2
dex higher metallicity in fully face-on galaxies than in fully edge-on
galaxies, for fixed mass.  For 15 of our GDDS galaxies we obtained
HST/ACS F814W images (Abraham et al., in preparation). One-third are
face-on, 1/3 are edge-on and 1/3 are intermediate (Figure~\ref{f6}).
Because this is a random effect, it would make the dispersion in the
metallicity larger, but any significant difference between the
$z\sim0.1$ and the $z\sim0.7$ samples would remain basically
unchanged.

\subsection{Different metallicity calibrators}\label{metcal}

A very important systematic effect to take into account is
linked to the several metallicity calibrators used by different
studies.  Kennicutt et al.\ (2003) showed that the metallicity
based on an electron-temperature sensitive method is on average a
factor of 2.5 lower than that derived using the empirical method by
Kewley \& Dopita (2002). This factor becomes $\sim2$, when compared
with the calibrator used by us (KK04).

We do not discuss here what the best method is, because this is still
controversial and is a strong function of what is observationally
available (e.g., what set of emission lines are used).  However,
because of these differences, we have paid particular attention to
comparing metallicities after converting to the {\it same calibrator},
namely, the KK04 one. This is the minimum requirement necessary to
make the detection of the redshift evolution of the M-Z relation
more robust.

\begin{figure}
\centerline{\epsfxsize=9cm \epsffile{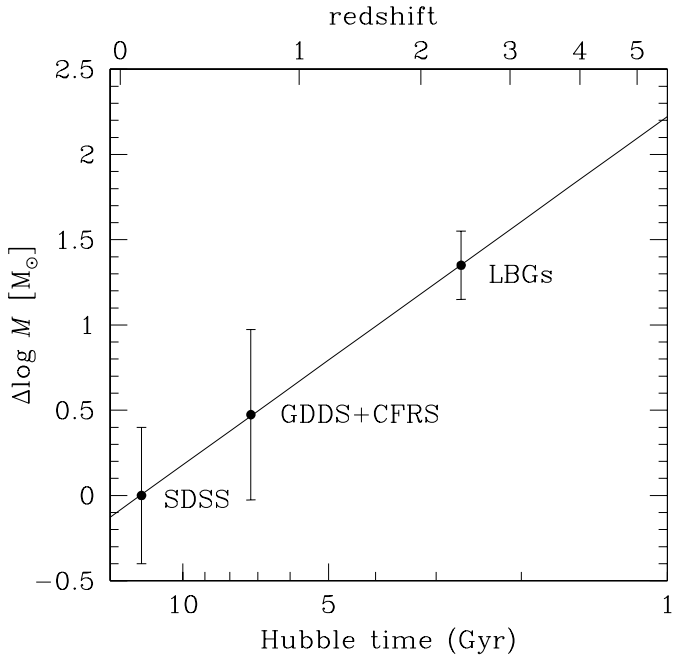}}
\caption {The points show the displacement in mass applied to the SDSS
M-Z relation (for which $\Delta \log M\equiv0$ by definition) to
match the GDDS+CFRS and LBG distributions. This describes how much
more massive a high-$z$ galaxy with a given metallicity would roughly
be with respect to local galaxies with similar metallicities.  The
error bars indicate the $\sim1\sigma$ mass {\it dispersion} at the
median metallicity in the samples, as seen in Figure~\ref{f13}.  The
small error bar for the LBGs is the effect of the small sample.  The
straight line is the linear correlation of the three points
(Eq.~\ref{dm}).}
\label{f14}
\end{figure}

\begin{figure*}
\centerline{\epsfxsize=16cm \epsffile{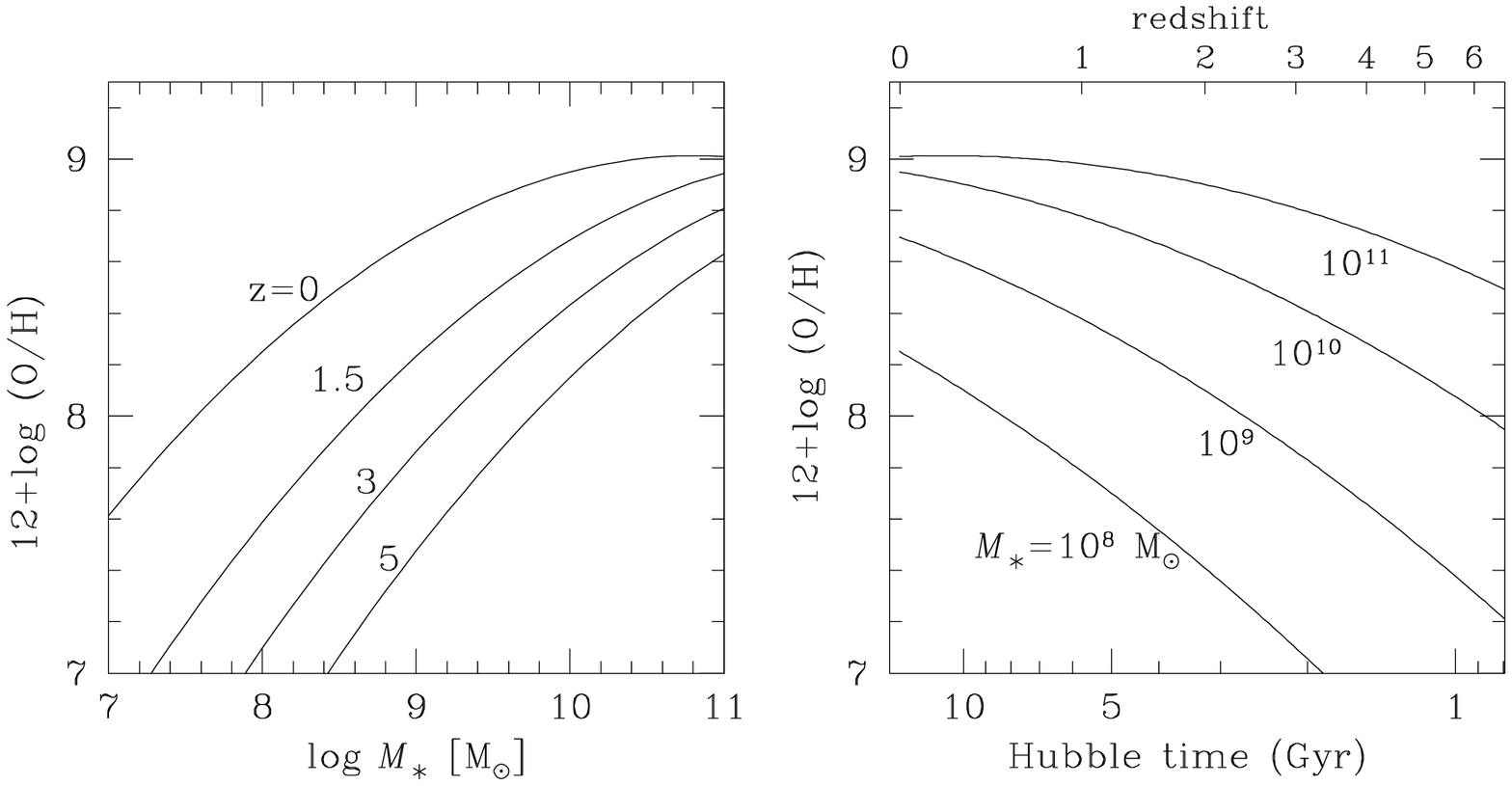}} 
\caption {$12+\log (\rm
O/H)$ as a function of stellar mass for constant redshifts ({\it left
panel}), or as a function of Hubble time for constant stellar masses
({\it right panel}). These relations are obtained by extrapolating at
all redshifts the $z\sim0.1$ SDSS and our $z\sim0.7$ M-Z relations.
We emphasize that this empirical model is derived using the KK04
metallicity calibrator and BG03 IMF, and is different if other
calibrators and IMFs are used.}
\label{f15}
\end{figure*}

\subsection{Dust extinction correction}

Another possible bias to consider is the dust extinction.  As stated
in \S\ref{dust}, dust extinction does not have a huge effect on the
metallicity. This decreases by only 0.2 dex if $A_V$ goes from 0 to
3. We applied $A_V=2.1$ and $A_V=1$ for GDDS and CFRS galaxies,
respectively. The choice of using a lower extinction in the CFRS is
justified by the higher luminosity of the CFRS galaxies, not entirely
accounted for by the relatively higher stellar masses of the same
sample with respect to the GDDS sample. Indeed, the CFRS is a $R$
selected survey that used the 4m Canada-France-Hawaii Telescope, thus
CFRS very likely could not identify highly extincted galaxies. The
GDDS is a $K$ selected survey that used a larger telescope, so suffers
less from this bias.

Brinchman et al.\ (2004) found that massive ($10^{10}$ M$_\odot
<M_\star<10^{11}$ M$_{\odot}$) star-forming galaxies on average suffer
3 times (in magnitudes) more extinction than low mass galaxies ($10^8$
M$_\odot < M_\star<10^9$ M$_\odot$), i.e., opposite to what we have
assumed.  Even if there was no bias in the CFRS sample, assuming a
larger extinction in the more massive galaxies would have a small
effect on the correlation, but the scatter would be larger.

\section{Modeling the mass-metallicity relation evolution}

\subsection{An empirical model}

We now attempt to derive an empirical model describing the redshift
evolution of the mass-metallicity relation.  This will allow to
predict the stellar mass (or the metallicity) of a galaxy at a given
redshift and metallicity (or stellar mass).

Let us arbitrarily assume that at high redshift the shape of the SDSS
relation is preserved, but it moves towards higher masses. We prefer
to move the SDSS relation in the $x$ direction instead of the $y$
direction, because this can better reproduce $z\sim0.7$ GDDS and CFRS
points in the M-Z plane. The best match is obtained if the M-Z
relation is moved to the right by $\Delta \log M \simeq 0.47$ (black
solid curve in Figure~\ref{f13}).  At higher redshift only the few
galaxies studied by Shapley et al.\ (2004) have measured mass and
metallicity, and these are represented by the rectangle in
Figure~\ref{f13}. We have corrected metallicities in order to take
into account systematic effects. Shapley et al.\ (2004) have used the
N2 calibrator, which saturates for high metallicities (Pettini \&
Pagel 2004).  If we again assume that the shape of the M-Z relation
is preserved, the curve intercepts the locus of the LBG points for
$\Delta \log M \simeq 1.35$.  This mass displacement, the one at
$z\sim0.7$, and the SDSS point for which $\Delta \log M\equiv0$, are
shown in the plane $t_H$ - $\Delta \log M$ ($t_H$ is the Hubble time)
of Figure~\ref{f14}. These three points are basically aligned along
the straight line (also shown)

\begin{equation}\label{dm}
\Delta \log M(t_H) = -2.0436\log t_H+2.2223
\end{equation}

\noindent
in the $\log-\log$ space.  If we combine Eq.\ \ref{dm} with Eq.\
\ref{newT}, we derive the general relation that gives the metallicity
of a galaxy at a given stellar mass and Hubble time:
 
\begin{eqnarray}
12 + \log  ({\rm O/H}) = -2.4412+2.1026 [\log M_\star+ \nonumber \\
-\Delta \log M(t_H)] -0.09649 [\log M_\star-\Delta \log M(t_H)]^2  \nonumber \\
\end{eqnarray}

\noindent
which is equivalent to:

\begin{eqnarray}\label{z}
12 + \log ({\rm O/H}) = -7.5903+2.5315 \log M_\star + \nonumber \\
-0.09649 \log^2 M_\star + 5.1733 \log t_H  -0.3944 \log^2 t_H + \nonumber \\
 -0.4030  \log t_H\log M_\star  \nonumber \\
\end{eqnarray}

\noindent
Eq.\ \ref{z} is shown in Figure~\ref{f15}, and predicts that low mass
galaxies have a steeper redshift evolution in metallicity than bigger
galaxies.  Of course we have to consider the big uncertainty due to
the way metallicities are measured.  However, Eq.~\ref{z} describes in
a simple and powerful fashion the evolution of the metal cosmic
abundance in the Universe for different mass bins. It predicts the
mean metallicity of galaxies at a given mass and redshift, but does
not give the time evolution of a single galaxy. This is done in the
next section.

\subsection{A closed-box model}\label{cb}

A qualitative explanation of the redshift evolution of the M-Z
relation (Figure~\ref{f15}) is provided by a SFR which stops earlier
in more massive galaxies than in low-mass galaxies (Heavens et
al.\ 2004; Juneau et al.\ 2005). Another way to describe this is a gas
fraction (gas mass over total mass) available for star formation, that
declines more rapidly in more massive galaxies, or a period of star
formation that lasts longer in less massive galaxies.

Theoretical scenarios explaining these observations, must properly
consider gas flows in both directions (in and out of the
galaxy). Galactic flows are mainly driven by the formation of massive
stars which blow out gas from the galaxy, and by the gravitational
potential of the galaxy (related to its mass), which on the one hand
confines the gas in the galaxy, and on the other attracts metal-poor
gas from the external IGM.  The chemical evolution of galaxies over
cosmic times was predicted by Pei \& Fall (1995) using infall, outflow
and closed-box (no flows) models. Larson (1974) described in his early
work that low mass galaxies are metal poor due to the gas loss after
supernova explosions. Ferrara \& Tolstoy (2000) found later on in
their models that indeed galaxies with gas mass $M_{gas}<10^9$ M$_\odot$
lose mass in outflows.

We tried to reproduce the M-Z relation evolution using the
closed-box approach. Although the large dispersion in the M-Z point
distribution can indicate that galactic flows play a role in the
chemical state of galaxies, the closed-box model allows one to start
to understand what basic assumptions are necessary to explain the
observations.  We use the P\'EGASE modeling (Fioc \& Rocca-Volmerange
1997, 1999) and assume that initially the galaxy is only made of gas
with a given total mass $M_{tot}$ and zero metallicity.  As the gas
collapses and stars form, the total mass remains constant, and part of
the gas is turned into stars.  Both stellar mass and metallicity
increase simultaneously as the galaxy evolves.  The other key
assumption is an exponential star formation SFR $\propto
\exp^{-t/\tau}$, with an e-folding time $\tau$ proportional to the
total baryonic mass of the galaxy, according to the following
empirical relation:

\begin{equation}\label{tau}
\log \tau = a \log M_{tot} +b
\end{equation}

\begin{figure}
\centerline{\epsfxsize=8.5cm \epsffile{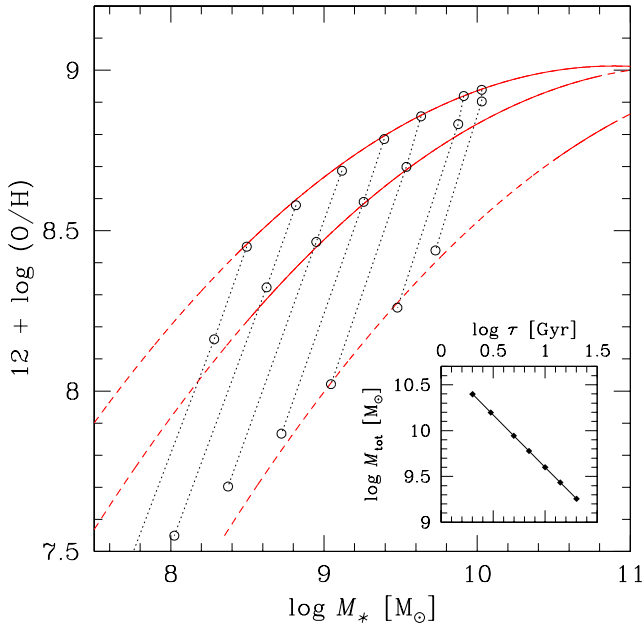}} 
\caption {Closed-box model predictions of $12+\log (\rm O/H)$ as a
function of stellar mass (dotted lines), for redshift of formation
$z_f=3.5$, and for increasing values of the initial mass (from left to
right). Results are not very sensitive to $z_f$, if this is in the
range $z_f=3-1000$. Open circles mark the three redshift epochs
$z=2.3$, 0.7 and 0.1, from bottom to top.  The dashed and solid lines
are results from observations at $z=2.3$, 0.7 and 0.1 from bottom to
top (as shown in Figure~\ref{f13}, see text). These lines are solid in
the mass range covered by observations. The small insert shows the
e-folding time as a function of the total galaxy mass adopted by the
model, also described by Eq.~\ref{tau} for $a=-0.95$ and $b=10.6$. The
mass intervals marked by the diamonds are as in the dotted lines of
the M-Z relation.}
\label{f16}
\end{figure}

\noindent
For $a<0$, the SFR declines more rapidly in more massive galaxies.  We
tune this equation to reproduce the proposed M-Z relation evolution
of Eq.~\ref{z} and Figure~\ref{f13}.  This occurs for $a=-0.88$ and
$b=9.42$ (small insert in Figure~\ref{f16}).  The results of the model,
assuming a redshift of formation $z_f=3.5$, are shown by the dotted
lines in Figure~\ref{f16}; they are not very sensitive to $z_f$, if this
is in the range $z_f=3-1000$. A galaxy with a stellar mass of
$10^{8.0}$ M$_\odot$ at $z=2.3$ will have a stellar mass of $10^{8.8}$
M$_\odot$ at $z=0.1$, i.e.\ a factor of 6 higher than 9.5 Gyr before
($z=2.3$). Another galaxy with stellar mass $10^{9.7}$ M$_\odot$ at
$z=2.3$, will only have doubled its stellar mass after the same time
interval has passed. The increase in metallicity is a factor of
10 in the less massive galaxy, and only a factor of 3 in the more
massive galaxy.  This simple model, where $\tau$ is a function of the
mass, provides a good explanation of the observed results. Without any
variation of $\tau$, it is not possible to account for the M-Z
relation at $z\sim0.1$, and all galaxies would have solar
metallicity, regardless of mass and redshift of formation (if in the
range $z_f=3-1000$).

\section{Discussion}

The mass-metallicity (or the luminosity-metallicity) relation is a
well known phenomenon identified in the local Universe in several
surveys (Faber 1973; Lequeuex et al.\ 1979; Garnett \& Shields 1987;
Skillman, Kennicutt, \& Hodge 1989; Zaritsky et al.\ 1994). Although
any correlation is more difficult to detect at high redshift, due to
the limited size of surveys, it is already clear that galaxies with a
given luminosity are more metal poor than similarly luminous galaxies
in the local Universe (Kobulnicky \& Koo 2000; Pettini et al.\ 2001;
Lilly et al.\ 2003; Liang et al.\ 2004; Shapley et al.\ 2004).  Only
recently it was possible to firmly establish that the relation between
mass and luminosity persists at high redshift ($z\sim0.7$; KK04),
thanks primarily to the large sample of galaxies available (about 200
from TKRS) and to the large interval in luminosity spanned. Such a
relation at high redshift is displaced with respect to the local
relation.

All this is generally based on the $B$ luminosity, a parameter that
poorly constrains the galaxy mass, because the $B$ luminosity is
easily extincted by dust, and represents short-lived massive stars
more than the bulk of the stellar mass.  Stellar mass is a more
meaningful galaxy parameter.

The stellar mass and metallicity of 56 galaxies in our sample
($0.4<z<1$) are correlated (at $\sim6\sigma$ significance level).
This relation is much stronger than the luminosity-metallicity
relation for the same sample (for which the dispersion around the best
fit is 65\% higher), or for the TKRS sample at similar
redshifts.

This M-Z relation is displaced towards higher stellar masses, and/or
lower metallicities, with respect to the $z\sim0.1$ M-Z relation for
SDSS star-forming galaxies.  The rate at which mass and metallicity
evolve with time can be estimated, also using results from the 7
$z\sim2.3$ LBGs studied by Shapley et al.\ (2004). Although the
$z\sim2.3$ sample is too small to establish any relation between mass
and metallicity, we used it to normalize a hypothetical M-Z
relation at $z\sim2.3$. Such an additional information is very important,
because it doubles the time interval: the Universe at $z\sim2.3$ is
4.6 Gyr younger than at $z\sim0.7$, which is 4.9 Gyr younger than at
$z\sim0.1$.  We assume that the shape of M-Z relation
found at $z\sim0.1$ does not change with time, and move it on the
stellar-mass axis, to find the best match with the observed points at
$z=0.7$ and 2.3.

We parametrize this displacement of mass and derive a general relation
for the metallicity that is a function of the stellar mass and of the
Hubble time (Eq.\ \ref{z} and Figure~\ref{f15}). With this generalized
relation, we can statistically predict the mass of a galaxy at a given
redshift and metallicity. Eq.~\ref{z} describes in a simple and
powerful fashion the evolution of the cosmic metal abundance as a
function of the galaxy stellar mass. It can also be used to test
predictions of semi-analytic models (Somerville, Primack \& Faber
2001), analytic models (Pei, Fall \& Hauser 1999), or numerical
simulations (Nagamine et al.\ 2004).

Any attempt to detect any cosmic chemical evolution would fail if this
is done combining galaxies with very different stellar masses.  Our
empirical model indicates that massive galaxies evolve in terms of
metallicity more slowly than less massive galaxies.  The DLAs offer
the opportunity to investigate consistently the chemical evolution for
a particular class of galaxies over a large fraction of the cosmic
time. It is now commonly accepted that DLAs do show chemical evolution
(Prochaska et al.\ 2003). This evolution can be compared with our
predictions to derive the stellar masses of DLA galaxies. Of course,
big uncertainties can lead to systematic errors. For instance, it is
hard to estimate any systematic difference between the metallicity
derived from absorption lines and emission lines. Moreover,
statistically, DLAs are tracing the outskirt of galaxies (Chen et al.\
2005) where metallicity is generally lower than the metallicity in the
central star-forming region (Ellison et al. 2005).  Chen et al.\
(2005) estimated a metallicity gradient of $0.041\pm0.012$ dex
kpc$^{-1}$.

\begin{figure}
\centerline{\epsfxsize=9cm \epsffile{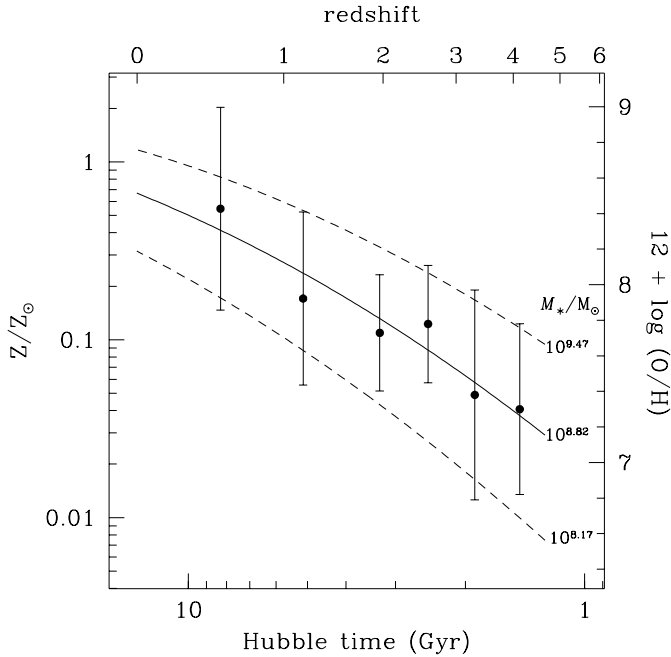}} 
\caption {Comparison
between mean metallicity in DLAs for different redshift bins (Savaglio
2001), and our empirical model of Eq.\ \ref{z}. The best fit gives a
stellar mass for DLAs of $10^{8.82}$ M$_\odot$ (solid line) with a
dispersion of 0.65 dex (dashed lines). In this figure we made a
correction of $-0.3$ dex to the model to account for
a metallicity gradient in DLAs (Chen et al.\ 2005). The best-fit mass
would be 0.3 dex higher if the $R_{23}$ metallicity, used to calibrate
the empirical model, is underestimated by a factor of 2 (Kennicutt et
al.\ 2003).}
\label{f17}
\end{figure}

We compared DLA observations to our empirical model, after applying to
the model a constant negative correction in metallicity of 0.3
dex. Such an assumption is justified by assuming a typical impact
parameter of the DLA from the galaxy center of 10 kpc, and considering
that our empirical relation is derived from galaxies observed over a
region that is on average about 7 kpc across.  We take the mean
metallicity in redshift bins derived from a sample of 87 DLAs after
dust depletion correction (Savaglio 2001). This is similar to the one
more recently derived by Prochaska et al.\ (2003) using 100 DLAs. The
result is shown in Figure~\ref{f17}. Remarkably, the best fit predicts
that the stellar mass of DLA galaxies is $M_\star = 10^{8.82\pm0.65}$
M$_\odot$, all the way from $z\sim0.5$ to 4.1. This suggests that
typical DLAs are not originating in very low-mass dwarf galaxies, but
are intermediate mass systems. The estimated stellar mass would be 2
times higher (+0.3 dex) if the metallicity derived using the $R_{23}$
parameter is overestimated by a factor of 2 (Kennicutt et al.\ 2003;
KK04).

Figure~\ref{f15} and Eq.~\ref{z} also suggests that local dwarf
star-forming galaxies, with metallicities of $\sim 1/10$ solar (Aloisi
et al.\ 2003; and references therein), have stellar masses of the
order of $10^7$ M$_\odot$. In the distant Universe, the star-forming
galaxies at $z\sim2.1$ of the K20 with stellar masses (after
correcting for BG03 IMF) of $\sim10^{11}$ M$_\odot$ (Fontana et al.\
2004) would have a mean metallicity slightly over solar, in agreement
with results derived using UV stellar features by de Mello et al.\
(2004). In a previous GDDS work (Savaglio et al.\ 2004) we have
studied the ISM absorption in a sample of 13 galaxies at mean redshift
$z\sim1.6$ and found indication of high metal enrichment. The average
stellar mass of this sample is $M_\star = 10^{10.4}$ M$_\odot$, which
leads again, according to Eq.~\ref{z}, to metallicity slightly above
solar.

For the LBGs studied by Pettini et al.\ (2001) at $z=3-3.4$, applying
the KK04 calibrator and a modest dust extinction of $A_V=1$, the
reported emission line fluxes of 4 galaxies give $12+\log {\rm (O/H)}
= 8.4-8.8$. At the redshift of the galaxies, Eq.~\ref{z} gives a mean
stellar mass of $M_\star = 10^{10.4}$ M$_\odot$, i.e.\ about a factor
of 2 less massive than the mean value in the $z\sim2.3$ LBGs of
Shapley et al.\ (2004).

Massive LBGs are often used to derive the high-$z$ SFR history of the
Universe (the Madau plot; Madau et al.\ 1996). The Madau plot is then
used to derive the cosmic chemical evolution. A discrepancy with the
DLA metallicity evolution (which is lower and steeper) was noticed by
Madau, Pozzetti, \& Dickinson (1998). The discrepancy is likely
because the metallicity evolution of different galaxies with
substantially different stellar masses are compared.  This is
confirmed by by Hopkins, Rao, \& Turnshek (2005), who found that DLAs
have a dominant role in the SFR density only in the late Universe
($z<0.6$), but at higher redshift their contribution is much less
important. We also note that even if the analytic model of the cosmic
metallicity evolution by Pei et al.\ (1999) resembles the DLA
metallicity evolution, this alone does not necessarily prove that the
cosmic chemical evolution is well represented by the DLA galaxies.

Our empirical model can be tested using recent results on a DLA sample
at $1.7<z<3.7$ (Ledoux et al.\ 2005) showing a correlation
between FeII absorption equivalent width (representing the galaxy
dynamical mass) and metallicity. Metal rich DLAs have larger FeII EW
than metal poor DLAs, a result which is reminiscent of the M-Z
relation. If properly translated, their correlation at a mean redshift
of $z\sim2.4$ is consistent with our M-Z relation (C.\ Ledoux,
private communication).

The observed evolution of the M-Z relation can be reproduced by a
closed-box scenario where the e-folding time is shorter for more
massive galaxies. A star formation more concentrated in time in
massive galaxies than in less massive galaxies is consistent with the
``downsizing'' scenario of galaxy formation. Other models cannot be
ruled out by our simple approach.

One shortcoming of our simple model is that it makes no attempt to consider
the impact of galaxy-galaxy mergers on the evolution. This is known to
evolve rapidly with redshift (Le F{\`e}vre et al.\ 2000). Mergers
have the most dramatic effect when they are between galaxies of
similar mass, and in many such cases the merger product will no longer be
star-forming and hence not be included in these samples. The existence
of a tight M-Z relation implies that an equal mass merger will shift
galaxies to the right in Figure~\ref{f16}, the net effect will be to
make the model tracks less steep. A model which included merging would
then have to have a different form for $\tau-M_{tot}$ in order to compensate
for this effect.  A full treatment of merging would require a detailed
semi-analytic (e.g., Springel et al.\ 2005) or hydrodynamical model
(Nagamine et al.\ 2004) of galaxy formation which is beyond the scope of
this work, though we believe the empirical relations we have
discovered will greatly inform such a comparison. We note that it is
much more straight-forward to compare stellar mass with metallicity as
galaxies assemble in a model as one has to make less assumptions
(e.g., about dust) than one does with a luminosity comparison.

\section{Summary}

In this paper, we investigated for the first time the high-$z$
mass-metallicity relation. We used galaxies from the GDDS and CFRS,
for which stellar mass estimates were possible thanks to the available
multi-band optical-NIR photometry (Crampton et al.\ 1995; Le F{\`e}vre et
al.\ 1995; Lilly et al.\ 1995; McCarthy et al. 2001; Chen et al.\
2002).

The metallicity was measured using the $R_{23}$ and $O_{32}$ over a
sample of star-forming galaxies at $0.4<z<1$. For the CFRS sample, we
recalculated values measured by Lilly et al.\ (2003) using the recent
calibrator provided by Kobulnicky \& Kewley (2004). Stellar mass and
metallicity are measured simultaneously for 56 galaxies.

Thanks to the large mass range, $M_\star=10^{8.2}-10^{10.8}$
M$_\odot$, and the size of the sample, we unambiguously detect a
$6\sigma$ significant correlation between stellar mass and metallicity
at $z\sim0.7$ (Eq.~\ref{met2}). Such a M-Z relation is displaced
with respect to the M-Z relation at $z\sim0.1$, towards higher
stellar masses, or lower metallicities. Galaxies at a given
metallicity are more massive than galaxies in the local universe with
similar metallicity. We attribute the origin of this to the redshift
evolution of the mass-metallicity relation. Although the shape of
the M-Z relation depends to some extend on the choise of the
metallicity calibrator, our result on the evolution of the M-Z
relation is robust, as all metallicities are estimated using the same
calibrator.

The stellar mass is more tightly correlated to metallicity than the
luminosity (the dispersion around the best fit is 1.6 times
smaller). Moreover, the redshift evolution of the M-Z relation is
more significant than the redshift evolution of the metallicity found
in large samples of galaxies with different stellar mass.

We find the M-Z evolution to be more rapid in lower-mass galaxies
indicating that they are still actively being constructed. In contrast the
more massive galaxies have already reached solar metallicity at
$z=1$ indicating that the bulk of their star-formation has completed.

Based on the observed results, we derive an empirical model for the
evolution of the mass-metallicity relation as a function of redshift
which works surprisingly well.  From this, one can estimate the
expected stellar mass of a galaxy at a given metallicity and
redshift. If our model is correct, we predict that the stellar mass of
DLAs is $M_\star = 10^{8.8\pm0.7}$ M$_\odot$ at any
redshift. According to the same model, the few LBGs at $\sim3.1$
studied by Pettini et al.\ (2001) may have stellar masses of the order
of $2\times10^{10}$ M$_\odot$.  The predicted masses are $\sim2$ times
higher if the $T_e$-based metallicity calibrator is used (Kennicutt et
al.\ 2003).

Our empirical model is nicely reproduced by a toy scenario of
galaxy formation where the important prescription is a star formation
that proceeds more rapidly in more massive galaxies, as described by
the downsizing scenario for galaxy formation. Any more detailed
model of galaxy formation must also reproduce the observed M-Z relation
as a function of redshift.

Future programs can test our predictions. For instance, by measuring
metallicities of galaxies with known masses in the $z\sim1.5$
Universe. Galaxies of the GDDS and K20 (Fontana et al.\ 2004;
Glazebrook et al.\ 2004) would be a suitable testbed to this goal. To
measure metallicities, NIR spectroscopy would be required, not an
easy, but possible, task for those faint galaxies.  Stellar masses of
DLAs can directly be measured by obtaining $K$ band photometry of QSO
fields. This can more efficiently be done using the NIR capabilities
of $HST$ over low-$z$ targets, to limit the confusion with the QSO
PSF.  Alternately ground-based Adaptive Optics may advance to the
point where this is possible with larger telescopes.  Finally, our
proposed model of star formation history of galaxies, which is a
function of mass, can be compared with the SFR history of the Universe
in mass bins (Juneau et al.\ 2004; Heavens et al.\ 2004).
 
\acknowledgments 

The authors thank the anonymous referee, Duilia de Mello, Lisa Kewley
and Christy Tremonti for invaluable comments.  C\'edric Ledoux \&
Christy Tremonti are acknowledged for providing results prior to
publication. This paper is based on observations obtained at the
Gemini Observatory, which is operated by the Association of
Universities for Research in Astronomy, Inc., under a cooperative
agreement with the NSF on behalf of the Gemini partnership: the
National Science Foundation (United States), the Particle Physics and
Astronomy Research Council (United Kingdom), the National Research
Council (Canada), CONICYT (Chile), the Australian Research Council
(Australia), CNPq (Brazil), and CONICET (Argentina).  K.G.\ and
S.S. acknowledge generous funding from the David and Lucille Packard
Foundation.  H.--W.C. acknowledges support by NASA through a Hubble
Fellowship grant HF-01147.01A from the Space Telescope Science
Institute, which is operated by the Association of Universities for
Research in Astronomy, Incorporated, under NASA contract NAS5-26555.

{}

\clearpage
\begin{landscape}
\tabletypesize{\tiny} 
\begin{deluxetable}{lccccccccccccccc} 
\tablecolumns{15} 
\tablecaption{GDDS galaxies}
\tablehead{ 
  \colhead{ID} & 
  \colhead{$z$} & 
  \colhead{$f_{\rm [OII]3727}$\tablenotemark{a}} & 
  \colhead{$f_{\rm H\gamma}$\tablenotemark{a,b}} &
  \colhead{$f_{\rm H\beta}$\tablenotemark{a,b}} &
  \colhead{$f_{\rm [OIII]4959}$\tablenotemark{a}} &
  \colhead{$f_{\rm [OIII]5007}$\tablenotemark{a}} &
  \colhead{$\log R_{23}$\tablenotemark{c}} &
   \colhead{$\log O_{32}$\tablenotemark{c}} &
  \colhead{$12+\log (\rm O/H)$\tablenotemark{d}} &
  \colhead{$V$\tablenotemark{e}} & 
  \colhead{$z'$ \tablenotemark{e}} & 
  \colhead{$K$\tablenotemark{e}} & 
  \colhead{$M_{B,AB}$} & 
  \colhead{$\log M_\star$\tablenotemark{f}} \\}
\startdata
SA02-0585& 0.826&$ 15.52\pm 0.69$ & $1.98\pm1.04$ &  $5.21\pm 1.40$ 	&$  3.05\pm 1.21$ 	&$ 16.78\pm 1.16$ 	 & $0.95\pm 0.12$ & $ -018\pm 0.06$ &$ 8.389\pm0.235$ & 23.76& 22.57 &$ >20.6$ &$ -19.96$ &$  8.98\pm 0.09$ \\ 
SA02-0756& 0.864&$ 11.03\pm 0.59$ & $1.33\pm0.49$ &  $2.69\pm 0.82$ 	&$  2.99\pm 1.11$ 	&$  8.97\pm 1.94$ 	 & $1.06\pm 0.14$ & $ -0.25\pm 0.10$ &$ 8.211\pm0.343$ & 24.75& 23.48 &$ >20.6$ &$ -19.10$ &$  8.81\pm 0.38$ \\ 
SA12-5685& 0.961&$ 16.47\pm 0.46$ & . . . 			& $10.72\pm 2.11$ 	& . . . 			&$ 20.29\pm 2.52$ 	&  $0.71\pm 0.09$ & $ -0.07\pm 0.06$ &$ 8.754\pm0.099$ & 23.82 & 22.84 &$ 20.12$ &$ -20.71$ &$  9.38\pm 0.13$ \\ 
SA12-5722& 0.842&$  6.35\pm 0.76$  & . . . 			&  $7.55\pm 2.02$ 	& . . . 			&$  5.32\pm 0.78$ 	 & $0.38\pm 0.12$ & $ -0.24\pm 0.08$ &$ 8.994\pm0.058$ & 24.52& 22.39 &$ 18.37$ &$ -20.43$ &$ 10.69\pm 0.10$ \\ 
SA12-7099& 0.567&$  6.97\pm 0.65$  & . . . 			& $12.60\pm 0.46$ 	& . . . 			& $<2.0$ 			 & $0.02\pm 0.06$ & $ -1.30 \pm0.65$ &$ >9.1$ & 23.01& 20.88 &$ 17.54$ &$ -20.09$ &$ 10.63\pm 0.08$ \\ 
SA12-7205& 0.567&$  6.31\pm 0.49$  & . . . 			&  $8.31\pm 0.32$ 	& . . . 			&$<1.4$ 			 & $0.20\pm 0.05$ & $ -0.82\pm 0.14$ &$ >9.1$ & 23.50& 22.10 &$ 19.13$ &$ -19.41$ &$  9.86\pm 0.25$ \\ 
SA12-7660& 0.791&$ 10.89\pm 0.40$ & $3.92\pm0.41$ & $10.85\pm 0.60$ 	& . . . 			&$  2.23\pm 0.89$ 	 & $0.32\pm 0.05$ & $ -0.85 \pm0.14$ &$ 9.017\pm0.020$ & 23.91& 22.01 &$ 18.53$ &$ -20.56$ &$ 10.48\pm 0.14$ \\ 
SA12-7939& 0.664&$  5.79\pm 0.31$  & $0.91\pm0.27$ &  $2.09\pm 0.30$ 	&$  2.33\pm 0.28$ 	&$  5.30\pm 0.50$ 	 & $0.92\pm 0.07$ & $ -0.17 \pm0.06$ &$ 8.438\pm0.125$ & 24.38& 23.23 &$ >20.6$ &$ -18.94$ &$  8.84\pm 0.05$ \\ 
SA12-8250& 0.766&$  5.80\pm 0.40$  & . . . 			&  $3.42\pm 0.47$ 	& . . . 			&$  2.35\pm 0.69$ 	 & $0.59\pm 0.07$ & $ -0.56\pm 0.11$ &$ 8.850\pm0.064$ & 24.57& $>23.5$ &$ 20.64$ &$ -19.39$ &$  9.55\pm 0.34$ \\ 
SA15-4272& 0.918&$ 10.08\pm 0.26$ & . . . 			&  $3.47\pm 1.41$ 	&$  4.93\pm 1.32$ 	&$ 17.55\pm 1.11$ 	 & $1.05\pm 0.18$ & $ 0.06 \pm0.06$ &$ 8.276\pm0.417$ & 24.96& 22.99 &$ >20.6$ &$ -19.32$ &$ 10.10\pm 0.31$ \\ 
SA15-4662& 0.895&$  8.83\pm 0.26$  & $2.68\pm0.25$ &  $8.35\pm 0.59$ 	& . . . 			&$  4.99\pm 0.98$ 	 & $0.42\pm 0.05$ & $ -0.41 \pm0.08$ &$ 8.969\pm0.025$ & 25.18& 23.25 &$ 20.47$ &$ -19.30$ &$  9.66\pm 0.29$ \\ 
SA15-5596& 0.890&$ 24.47\pm 0.37$ & $2.71\pm0.35$ &  $7.84\pm 0.67$ 	& . . . 			&$ 16.12\pm 1.87$ 	 & $0.91\pm 0.05$ & $ -0.34\pm 0.06$ &$ 8.443\pm0.087$ & 24.18& 23.07 &$ >20.6$ &$ -19.66$ &$  8.94\pm 0.29$ \\ 
SA15-6565& 0.955&$ 23.78\pm 0.53$ & $3.86\pm0.49$ & $13.50\pm 1.94$ 	&$ 10.97\pm 2.19$ 	&$ 34.66\pm 3.25$ 	 & $0.80\pm 0.07$ & $ -0.01\pm 0.06$ &$ 8.646\pm0.092$ & 24.68& 23.29 &$ >20.6$ &$ -19.13$ &$  9.81\pm 0.47$ \\ 
SA15-7399& 0.621&$  1.79\pm 0.33$  & . . . 			&  $<1.5$         		&$  1.97\pm 0.30$ 	&$  4.43\pm 0.32$ 	 & $0.79\pm 0.16$ & $ 0.26 \pm0.10$ &$ <8.7$ & $>26.3$ & $>23.5$  &$ >20.6$ &$ -17.34$ & . . . \\ 
SA22-0040& 0.818&$ 22.74\pm 0.45$ & . . . 			&  $8.82\pm 1.60$ 	&$  9.89\pm 1.11$ 	&$ 26.61\pm 1.06$ 	 & $0.93\pm 0.08$ & $ -0.08 \pm0.05$ &$ 8.436\pm0.150$ & 23.78& 22.60 &$ 20.28$ &$ -19.80$ &$  9.36\pm 0.40$ \\ 
SA22-0145& 0.753&$ 19.29\pm 0.56$ & . . .			&  $8.81\pm 0.74$ 	&$  7.91\pm 0.79$ 	&$ 17.45\pm 1.52$ 	 & $0.82\pm 0.04$ & $ -0.17 \pm0.05$ &$ 8.607\pm0.065$ & 23.72& 22.42 &$ 20.19$ &$ -19.50$ &$  9.88\pm 0.24$ \\ 
SA22-0563& 0.786&$ 32.69\pm 0.41$ & $8.38\pm0.69$ & $17.63\pm 0.71$ 	&$  9.28\pm 1.15$ 	&$ 15.79\pm 1.23$ 	 & $0.67\pm 0.03$ & $ -0.40 \pm0.05$ &$ 8.782\pm0.035$ & 23.17& 22.04 &$ 19.71$ &$ -20.42$ &$  9.58\pm 0.21$ \\ 
SA22-0619& 0.671&$  3.64\pm 0.20$  & . . . 			&  $1.89\pm 0.17$ 	& . . . 			&$  1.89\pm 0.27$ 	 & $0.68\pm 0.05$ & $ -0.45 \pm0.07$ &$ 8.774\pm0.057$ & 24.82& 22.77 &$ 19.37$ &$ -18.74$ &$ 10.08\pm 0.27$ \\ 
SA22-0630& 0.753&$  4.81\pm 0.42$ & $4.88\pm0.36$ & $14.50\pm 0.66$ 	&$  2.74\pm 0.61$ 	&$  7.83\pm 1.79$ 	 & $0.11\pm 0.05$ & $ 0.05 \pm0.10$ &$ 9.086\pm0.012$ & 23.61& 21.63 &$ 19.07$ &$ -20.26$ &$  9.84\pm 0.26$ \\ 
SA22-0643& 0.787&$  7.63\pm 0.39$ & . . . 			&  $2.91\pm 0.58$ 	& . . . 			&$  2.37\pm 0.98$ 	 & $0.76\pm 0.10$ & $ -0.67\pm 0.15$ &$ 8.664\pm0.131$ & 23.95& 22.70 &$ 20.28$ &$ -19.65$ &$  9.40\pm 0.35$ \\ 
SA22-0751& 0.471&$ 20.69\pm 0.70$ & . . . 			&  $7.60\pm 0.33$ 	&$  3.83\pm 0.42$ 	&$ 11.15\pm 0.46$ 	& $0.83\pm 0.04$ & $ -0.43 \pm0.05$ &$ 8.578\pm0.054$ & 23.31& 22.12 &$ 20.42$ &$ -18.62$ &$  9.04\pm 0.26$ \\ 
SA22-0926& 0.785&$  8.55\pm 0.51$ & . . . 			&  $4.61\pm 1.04$ 	&$  4.91\pm 1.15$ 	&$  9.03\pm 1.54$ 	 & $0.79\pm 0.10$ & $ -0.08 \pm0.08$ &$ 8.657\pm0.140$ &  24.07& 22.65 &$ >20.6$ &$ -19.45$ &$  9.48\pm 0.53$ \\ 
SA22-0997& 0.642&$  8.33\pm 0.52$ & . . . 			&  $2.12\pm 0.66$ 	&$  2.11\pm 0.43$ 	&$  7.49\pm 0.37$ 	 & $1.05\pm 0.14$ & $ -0.23 \pm0.06$ &$ 8.232\pm0.339$ & 24.71& 22.97 &$ >20.6$ &$ -18.12$ &$  9.70\pm 0.13$ \\ 
SA22-1534& 0.469&$  2.84\pm 0.57$ & . . . 			& $<0.9$          		& . . . 			&$  1.95\pm 0.39$ 	 & $0.94\pm 0.19$ & $ -0.33 \pm0.12$ &$< 8.4$ &  24.54& 23.03 &$ >20.6$ &$ -17.31$ &$  8.37\pm 0.47$ \\ 
SA22-1674& 0.879&$ 19.36\pm 0.37$ & $3.32\pm0.35$ &  $9.33\pm 0.65$ 	&$ 10.64\pm 1.59$ 	&$ 30.31\pm 1.21$ 	 & $0.89\pm 0.04$ & $ 0.04 \pm0.05$ &$ 8.510\pm0.061$ & 24.52& $>23.5$ &$ >20.6$ &$ -19.24$ &$  9.69\pm 0.57$ \\ 
SA22-2196& 0.627&$  8.98\pm 0.52$ & . . . 			&  $2.75\pm 0.74$ 	&$  3.73\pm 0.52$ 	&$  8.57\pm 0.49$ 	 & $1.00\pm 0.12$ & $ -0.15 \pm0.06$ &$ 8.335\pm0.261$ & 24.75& $>23.5$ &$ >20.6$ &$ -18.28$ &$  9.27\pm 0.65$ \\ 
SA22-2491& 0.470&$  7.23\pm 0.65$ & . . . 			&  $3.51\pm 0.30$ 	&$  3.40\pm 0.41$ 	&$  9.32\pm 0.48$    & $0.85\pm 0.05$ & $ -.04 \pm0.06$ &$ 8.571\pm0.073$ & 24.05& 22.67 &$ >20.6$ &$ -17.78$ &$  8.19\pm 0.44$ \\ 
SA22-2541& 0.617&$  8.01\pm 0.48$ & . . . 			&  $4.83\pm 0.70$ 	&$  1.33\pm 0.54$ 	&$  3.77\pm 0.43$ 	& $0.60\pm 0.07$ & $ -0.49\pm 0.08$ &$ 8.846\pm0.064$ & 23.33 & 21.52 &$ 18.78$ &$ -19.75$ &$  9.95\pm 0.42$ \\ 
\enddata
\tablenotetext{a}{Fluxes, in $10^{-18}$ erg s$^{-1}$ \cm, are not corrected for dust extinction.}
\tablenotetext{b}{Corrected for Balmer stellar absorption.}
\tablenotetext{c}{Corrected for dust extinction assuming $A_V=2.1$.}
\tablenotetext{d}{Errors do not include systematic uncertainties.}
\tablenotetext{e}{Vega magnitudes from LCIRS.}
\tablenotetext{f}{Stellar masses in units of solar masses.}
\end{deluxetable}
\clearpage
\end{landscape}

\begin{table}
\caption[t1]{GDDS composite emission line  fluxes}
\begin{center} 
\begin{tabular}{lcccccc} 
\tableline\tableline&&&&&&\\[-5pt] 
Line ratio & Measured & \multicolumn{2}{c}{Theory \tablenotemark{a}}  && \multicolumn{2}{c}{$A_V$\tablenotemark{b}} \\
[4pt]\cline{3-4}\cline{6-7}\\[-4pt] 
     &          & $10^4$ K & 5000 K && $10^4$ K & 5000 K \\
[5pt]\tableline&&&\\[-5pt] 
H$\gamma$/H$\beta$ & $0.292\pm0.026$          & 0.470 & 0.458 & & $3.47\pm0.64$ & $3.28\pm0.64$\\
H$\delta$/H$\beta$ & $0.180\pm0.017$          & 0.262 & 0.250 & & $1.91\pm0.48$ & $1.67\pm0.48$ \\
H$\epsilon$/H$\beta$ & $0.127\pm0.018$        & 0.159 & 0.153 & & $1.00\pm0.61$ & $0.83\pm0.61$ \\
H8/H$\beta$    & $0.035\pm0.016$        & 0.107 & 0.102 & & $4.57\pm1.89$ & $4.37\pm1.89$ \\
$\rm [OII]$/H$\beta$ & $2.280\pm0.081$        & . . . & . . . & & . . . &. . .\\
$\rm [OIII]\lambda5007$/H$\beta$ & $2.262\pm0.088$ &  . . . & . .. &&. . .&. . .\\
$\rm [OIII]\lambda4959$/H$\beta$ & $0.647\pm0.042$ & . . . & . . .  &&. . .&. . .\\
$\rm [OIII]\lambda4363$/H$\beta$ & $<0.03$ & . . .& . . . &&. . .&. . .\\
$\rm [NeIII]$/H$\beta$ & $0.208\pm0.024$ &  . . .  & . . . &&. . .&. . .\\
[2pt]\tableline
& & & Weighted mean & & $2.13\pm0.32$ & $1.92\pm0.32$ \\
[2pt]\tableline
\end{tabular}
\tablenotetext{a}{As expected from atomic physics, for two
temperatures, and assuming no dust extinction.}
\tablenotetext{b}{Visual extinction, given the observed flux ratio and
assuming two temperatures.}
\end{center}
\end{table}

\begin{table}
\scriptsize
\caption[t1]{CFRS sample}
\begin{center} 
\begin{tabular}{lcccccc} 
\tableline\tableline&&&&&&\\[-5pt] 
ID  & $z$ & $K$ & $M_{B,AB}$\tablenotemark{a} & \multicolumn{2}{c}{$12+\log (\rm O/H)$} & $\log M_\star$ \\
[4pt]\cline{5-6}\\[-4pt] 
& &(mag) & (mag)& LCS & This paper\tablenotemark{b} & [M$_\odot$]\\
[5pt]\tableline&&&&&&\\[-5pt] 
03.0062 & 0.826 & $ 20.17\pm 0.13$ & $-22.30$ &  9.10 & $ >9.06 $              & $ 10.57\pm 0.19$ \\ 
03.0085 & 0.609 & $ 21.32\pm 0.39$ & $-20.40$ &  8.84 & $  8.88\pm 0.04$ & $  9.62\pm 0.35$ \\ 
03.0125 & 0.789 & $ 20.89\pm 0.21$ & $-21.02$ &  8.94 & $  8.96\pm 0.05$ & $ 10.21\pm 0.19$ \\ 
03.0145 & 0.603 & $ 21.20\pm 0.28$ & $-20.87$ &  8.74 & $  8.77\pm 0.06$ & $  9.63\pm 0.20$ \\ 
03.0261 & 0.697 & $ 20.18\pm 0.11$ & $-20.95$ &  8.74 & $  8.76\pm 0.06$ & $ 10.46\pm 0.14$ \\ 
03.0327 & 0.609 & $ 20.71\pm 0.22$ & $-20.49$ &  8.92 & $  8.95\pm 0.07$ & $  9.99\pm 0.27$ \\ 
03.0488 & 0.605 & $ 21.31\pm 0.38$ & $-20.89$ &  8.64 & $  8.69\pm 0.07$ & $  9.44\pm 0.28$ \\ 
03.0570 & 0.646 & $ 21.79\pm 0.60$ & $-20.51$ &  8.64 & $  8.68\pm 0.10$ & $  9.42\pm 0.57$ \\ 
03.0595 & 0.605 & $ 20.65\pm 0.21$ & $-20.92$ &  8.90 & $  8.94\pm 0.05$ & $  9.99\pm 0.19$ \\ 
03.0599 & 0.479 & $ 21.25\pm 0.36$ & $-20.66$ &  8.30 & $  8.39\pm 0.22$ & $  9.54\pm 0.25$ \\ 
03.0879 & 0.601 & . . .            & $-19.94$ &  8.90 & $  8.94\pm 0.04$ &  . . .  \\ 
03.0999 & 0.706 & $ 19.93\pm 0.11$ & $-21.57$ &  8.94 & $ >8.94 $                 & $ 10.51\pm 0.20$ \\ 
03.1016 & 0.702 & $ 21.29\pm 0.38$ & $-20.49$ &  8.80 & $  8.84\pm 0.04$ & $  9.97\pm 0.46$ \\ 
03.1112 & 0.768 & . . .            & $-21.00$ &  8.56 & $  8.61\pm 0.10$ &    . . .  \\
03.1138 & 0.768 & $ 21.61\pm 0.51$ & $-20.70$ &  8.56 & $  8.63\pm 0.08$ &    . . .  \\
03.1309 & 0.617 & $ 19.32\pm 0.05$ & $-21.86$ &  8.94 & $  8.94\pm 0.06$ & $ 10.82\pm 0.14$ \\ 
03.1349 & 0.617 & $ 19.29\pm 0.05$ & $-21.57$ &  8.98 & $  9.01\pm 0.03$ & $ 10.48\pm 0.35$ \\ 
03.1367 & 0.703 & . . .            & $-20.78$ &  8.80 & $  8.84\pm 0.05$ &    . . .  \\
03.1375 & 0.635 & $ 23.29\pm 2.37$ & $-20.43$ &  8.84 & $  8.88\pm 0.07$ &    . . .  \\
03.1534 & 0.794 & . . .            & $-20.75$ &  8.58 & $  8.63\pm 0.05$ &   . . .  \\
03.9003 & 0.618 & . . .            & $-21.67$ &  9.02 & $  9.03\pm 0.03$ &   . . .  \\
10.0478 & 0.752 & . . .            & $-21.42$ &  8.58 & $  8.64\pm 0.11$ &   . . .  \\
10.1116 & 0.709 & . . .            & $-21.10$ &  8.96 & $ >8.95  $                &  . . .  \\ 
10.1213 & 0.815 & $ 20.63\pm 0.16$ & $-21.18$ &  8.86 & $  8.90\pm 0.16$ & $ 10.22\pm 0.31$ \\ 
10.1608 & 0.729 & $ 20.92\pm 0.22$ & $-20.67$ &  8.72 & . . .                         & $ 10.03\pm 0.28$ \\ 
10.1925 & 0.783 & $ 20.98\pm 0.16$ & $-21.17$ &  8.82 & $  8.87\pm 0.06$ & $  9.87\pm 0.31$ \\ 
10.2183 & 0.910 & $ 20.69\pm 0.12$ & $-22.14$ &  8.92 & $  8.95\pm 0.03$ & $ 10.30\pm 0.20$ \\ 
10.2284 & 0.773 & $ 21.46\pm 0.25$ & $-20.78$ &  8.76 & $ >8.61$               & $  9.85\pm 0.34$ \\ 
10.2418 & 0.796 & $ 20.14\pm 0.07$ & $-22.23$ &  9.02 & $ >9.00$               & $ 10.56\pm 0.14$ \\ 
10.2428 & 0.872 & $ 21.61\pm 0.29$ & $-21.49$ &  8.50 & $  8.54\pm 0.07$ & $  9.89\pm 0.16$ \\ 
10.2519 & 0.718 & $ 22.00\pm 0.41$ & $-20.07$ &  8.90 &   . . .                      & $  9.67\pm 0.21$ \\ 
10.2548 & 0.770 & $ 20.24\pm 0.08$ & $-21.32$ &  8.78 & $ <8.83  $                   & $ 10.56\pm 0.12$ \\ 
14.0072 & 0.621 & $ 21.30\pm 0.54$ & $-20.05$ &  8.58 & $  8.61\pm 0.07$ & $  9.55\pm 0.41$ \\ 
14.0129 & 0.903 & $ 20.63\pm 0.29$ & $-21.16$ &  8.92 & $ >8.90$             & $ 10.53\pm 0.29$ \\ 
14.0217 & 0.721 & . . .            & $-21.14$ &  8.70 & $  8.73\pm 0.08$ &  . . .  \\ 
14.0272 & 0.670 & $ 19.12\pm 0.07$ & $-22.16$ &  9.10 & $  9.08\pm 0.03$ & $ 10.52\pm 0.31$ \\ 
14.0393 & 0.603 & $ 19.76\pm 0.07$ & $-22.00$ &  8.86 & $  8.90\pm 0.06$ & $ 10.22\pm 0.15$ \\ 
14.0497 & 0.800 & $ 20.59\pm 0.14$ & $-21.45$ &  8.76 & $  8.82\pm 0.10$ & $ 10.05\pm 0.40$ \\ 
14.0538 & 0.810 & . . .            & $-21.42$ &  8.76 & $  8.79\pm 0.03$ &  . . .  \\ 
14.0605 & 0.837 & . . .            & $-20.94$ &  8.78 & $  8.81\pm 0.05$ &   . . .  \\
14.0725 & 0.580 & $ 20.39\pm 0.23$ & $-19.92$ &  9.00 & $  9.02\pm 0.03$ & $ 10.13\pm 0.25$ \\ 
14.0779 & 0.580 & . . .            & $-20.28$ &  8.98 & $  9.00\pm 0.05$ &  . . .  \\ 
14.0818 & 0.901 & . . .            & $-22.54$ &  8.86 & $  8.89\pm 0.08$ &  . . .  \\ 
14.0848 & 0.664 & $ 26.07\pm 9.99$ & $-20.37$ &  8.68 & $  8.73\pm 0.12$ &  . . .  \\ 
14.0972 & 0.677 & $ 20.34\pm 0.22$ & $-21.56$ &  8.64 & $  8.69\pm 0.06$ & $  9.92\pm 0.36$ \\ 
14.0985 & 0.809 & $ 21.20\pm 0.25$ & $-20.76$ &  8.92 & $  8.94\pm 0.05$ & $  9.97\pm 0.26$ \\ 
14.1087 & 0.659 & $ 21.20\pm 0.25$ & $-20.63$ &  8.70 & $  8.72\pm 0.08$ & $  9.53\pm 0.35$ \\ 
14.1126 & 0.746 & . . .            & $-20.71$ &  8.30 & $  8.53\pm 0.14$ &  . . .  \\ 
14.1189 & 0.753 & $ 21.04\pm 0.22$ & $-20.86$ &  8.88 & $ >8.81$              & $  9.84\pm 0.32$ \\ 
14.1190 & 0.754 & . . .            & $-21.99$ &  9.08 & $ >9.05$             &  . . .  \\ 
14.1258 & 0.647 & $ 21.51\pm 0.34$ & $-20.31$ &  8.82 & $  8.87\pm 0.03$ & $  9.59\pm 0.28$ \\ 
14.1386 & 0.744 & $ 19.90\pm 0.08$ & $-21.67$ &  8.98 & $  9.01\pm 0.04$ & $ 10.21\pm 0.44$ \\ 
14.1466 & 0.674 & $ 23.40\pm 3.70$ & $-20.36$ &  8.30 & $  8.57\pm 0.12$ &  . . .  \\ 
[2pt]\tableline
\end{tabular}
\end{center}
\end{table}

\setcounter{table}{2}
\begin{table}
\scriptsize
\caption[t1]{-- {\it Continued}}
\begin{center} 
\begin{tabular}{lcccccc} 
\tableline\tableline&&&&&&\\[-5pt] 
ID  & $z$ & $K$ & $M_{B,AB}$\tablenotemark{a} & \multicolumn{2}{c}{$12+\log (\rm O/H)$} & $\log M_\star$ \\
[4pt]\cline{5-6}\\[-4pt] 
& &(mag) & (mag)& LCS & This paper\tablenotemark{b} & [M$_\odot$]\\
[5pt]\tableline&&&&&&\\[-5pt] 
14.9705 & 0.609 & . . .            & $-21.22$ &  8.78 & $  8.83\pm 0.10$ &  . . .  \\ 
22.0274 & 0.504 & $ 19.90\pm 0.10$ & $-21.60$ &  8.52 & $  8.53\pm 0.11$ & $  9.88\pm 0.39$ \\ 
22.0322 & 0.915 & $ 21.12\pm 0.31$ & $-21.80$ &  8.30 & $  8.35\pm 0.22$ & $  9.93\pm 0.39$ \\ 
22.0417 & 0.593 & $ 21.22\pm 0.34$ & $-20.30$ &  8.80 & $  8.86\pm 0.08$ & $  9.73\pm 0.35$ \\ 
22.0429 & 0.624 & $ 20.49\pm 0.17$ & $-20.48$ &  8.76 & $  8.80\pm 0.09$ & $ 10.07\pm 0.20$ \\ 
22.0576 & 0.887 & . . .            & $-21.24$ &  8.70 & $  8.72\pm 0.04$ &  . . .  \\ 
22.0599 & 0.886 & . . .            & $-21.78$ &  8.76 & $  8.80\pm 0.04$ &  . . .  \\ 
22.0770 & 0.816 & $ 22.29\pm 0.90$ & $-21.47$ &  8.30 & $  8.60\pm 0.12$ &  . . .  \\ 
22.0919 & 0.472 & . . .            & $-20.22$ &  8.30 & $  8.41\pm 0.04$ &  . . .  \\ 
22.1119 & 0.514 & . . .            & $-21.87$ &  8.86 & $  8.90\pm 0.08$ &  . . .  \\ 
22.1313 & 0.817 & $ 21.16\pm 0.32$ & $-21.54$ &  8.54 & $  8.61\pm 0.14$ & $  9.92\pm 0.45$ \\ 
22.1350 & 0.510 & $ 21.50\pm 0.44$ & $-19.88$ &  8.80 & $  8.84\pm 0.06$ & $  9.55\pm 0.40$ \\ 
22.1528 & 0.665 & $ 21.84\pm 0.60$ & $-20.54$ &  8.60 & $  8.65\pm 0.06$ & $  9.39\pm 0.41$ \\ 
[2pt]\tableline
\end{tabular}
\tablenotetext{a}{As reported in LCS.}
\tablenotetext{b}{Errors do not include systematic uncertainties.}
\end{center}
\end{table}

\end{document}